\documentclass[letterpaper,twocolumn,10pt]{article}
\usepackage{usenix}
\usepackage{xurl}

\usepackage{amsmath, amssymb}
\usepackage{booktabs}
\usepackage{caption}
\usepackage{subcaption}
\usepackage{colortbl}
\usepackage{multirow}
\usepackage{tikz}
\usepackage{enumitem}
\usepackage{siunitx}
\usepackage[linesnumbered,ruled,vlined]{algorithm2e}
\usepackage[all]{nowidow}
\usepackage{diagbox, eqparbox, hhline}
\usepackage{fontawesome}
\usepackage{xcolor} 
\usepackage[numbers,sort]{natbib}
\usepackage{tcolorbox}
\usepackage{hyperref}
\usepackage{bbold}
\usepackage{graphicx}
\tcbuselibrary{breakable} 

\newlist{rqlist}{enumerate}{3}
\setlist[rqlist]{label=\textbf{RQ\arabic*}.,before=\raggedright,leftmargin=40pt,ref=RQ\arabic*}

\newif\ifshowtodos

\showtodostrue

\newtcolorbox{takeaway}{
    breakable,
    colback=gray!10,       
    colframe=black,        
    boxrule=0.8pt,         
    sharp corners,         
    left=3mm,              
    right=3mm,
    top=2mm,
    bottom=2mm,
    fontupper=\small\itshape
}
\newcommand{\resultline}[1]{%
\noindent\textbf{Result.} \textit{#1}%
}

\newtcolorbox{goal}{
    breakable,
    colback=white,       
    colframe=black,        
    boxrule=0.8pt,         
    sharp corners,         
    left=3mm,              
    right=3mm,
    top=2mm,
    bottom=2mm,
    fontupper=\small
}

\begin{document}

\date{}

\title{\Large \bf The Role of Learning in Evading ML-based Network Intrusion Detection}

\author{
{\rm Kyle Domico}\\
University of Wisconsin-Madison
\and
{\rm Jean-Charles Noirot Ferrand}\\
University of Wisconsin-Madison
\and
{\rm Patrick McDaniel}\\
University of Wisconsin-Madison
} 

\maketitle

\begin{abstract}
\noindent Machine Learning-based Network Intrusion Detection Systems (ML-NIDS) can be bypassed by rudimentary adversarial perturbations. Recent work has focused on identifying where such perturbations can realistically be applied by a host-side adversary. Yet every one of these attacks produces perturbations the same way: searching from scratch for every flow. The cost of an attack therefore grows in lockstep with the number of flows it must perturb, and real networks produce them by the tens of millions. In this paper, we show that using reinforcement learning to train lightweight perturbation-generating policies lets an adversary amortize that cost across flows it perturbs. Counting every detector query and every second an attack spends, training included, we compare learned policies against gradient, query-based, and random search across six ML-NIDS environments at two operating points, under both evasion and alert inflation. One successful adversarial example costs a learned policy 1.5 to 18 detector queries against 52 to 1,100 for the strongest search baseline, and the policy amortizes its training cost after 76 to 1,622 examples, a volume a monitored link produces in seconds of traffic. We further find that the RL formulation literature adopts by default is unnecessary for evasion, that the policy conditions on the flow it is given rather than converging on a fixed perturbation, and that it transfers to detectors and traffic it never trained against. The value of learning to attack ML-NIDS is therefore not a matter of effectiveness, but of scale.
\end{abstract}

\section{Introduction}\label{sec:intro}
Machine learning (ML) is used in contemporary network intrusion detection systems (NIDS). Real world evidence shows that ML-NIDS are actually deployed in security operations centers~\cite{crowley_sans_2025, cloudflare_one_2020, mink_everybodys_2023}. Accordingly, many papers in the last decade have asked if the vulnerabilities of ML transcend into ML-NIDS~\cite{apruzzese_evading_2018, sheatsley_robustness_2021, wang_manda_2021, liu_hard-label_2025, verkerken_what_2026}, and the answers have been unsettling: perturbations in the fractions of a percent to traffic flow features (e.g., NetFlow~\cite{claise_cisco_2004}) are enough to circumvent the entire system~\cite{apruzzese_evading_2018, sheatsley_adversarial_2022, sheatsley_robustness_2021}.

Much of the field's recent progress has come from interrogating \textit{where} those perturbations live. Seminal work established that real adversaries act in the problem space rather than the feature space~\cite{pierazzi_intriguing_2020}, that they interact with ML systems rather than ML models~\cite{apruzzese_real_2022}, and that unconstrained feature manipulation yields adversarial examples no adversary could physically produce~\cite{sheatsley_robustness_2021, verkerken_what_2026}. Realizability is the right problem: it is what makes adversarial examples in ML-NIDS a credible security phenomenon rather than academic exercise. One component, however, has gone unchallenged: the algorithm producing the perturbation. In feature space or problem space, under derived constraints or arbitrary ones, generation remains \textit{search-based}: gradient-guided optimization~\cite{diallo_sabre_2024, bhusal_pasa_2024, sheatsley_robustness_2021} or black-box query-based search~\cite{liu_hard-label_2025, nasr_defeating_2021, han_evaluating_2021}. Such procedures are deliberately data-agnostic: a per-flow search retains nothing between flows, so the millionth flow costs what the first did. Against benchmarks of tens of millions of flow records~\cite{moustafa_unsw-nb15_2015, sharafaldin_toward_2018}, fed by a network emitting them continuously ($\geq$10{,}000 pps~\cite{edwards_laymans_2025}), the cost of generation counted in detector queries stops being a footnote and becomes the feasibility question.

Reinforcement learning (RL) offers a different solution: rather than search independently for each perturbation, an adversary can learn a reusable policy for exploiting the ML-NIDS across flows. Training incurs a large one-time cost, after which the policy generates perturbations at low marginal cost. Such policies require neither gradients nor access to model internals and have been studied in image classification~\cite{domico_adversarial_2026}, malware detection~\cite{anderson_evading_2017, tsingenopoulos_how_2024}, LLM jailbreaking~\cite{chen_when_2024}, and ML-NIDS itself~\cite{hore_deep_2025, liu_amoeba_2023, apruzzese_deep_2020}. Throughout this paper, \textit{learning} therefore means \textit{policy learning}: training a perturbation-generating policy through interactions with an ML-NIDS and reusing it on previously unseen flows.

Yet learned policies have not replaced search. The obstacle is not effectiveness: they can generate effective adversarial examples. The unresolved question is cost: \textit{why spend hours training a policy when search can produce an adversarial example now?} Answering it requires measuring how many adversarial examples must be generated before training pays for itself. This has not been measured, leaving the case for RL to rest on post-training effectiveness alone. \textbf{We argue that the role of learning in evading ML-NIDS is settled not by whether the policy evades detection, but by when it becomes cheaper than the search it replaces.}

We answer this by treating adversarial example generation against ML-NIDS as an \textit{amortization} problem. A learned policy and a per-input search buy the same product at different prices: search charges a fixed rate per example, while a policy charges upfront for training and a discount thereafter. The two cross at a break-even volume $N^{*}$, the number of adversarial examples beyond which learning is cheaper, which we measure by counting every detector query either method consumes, training included~\cite{arp_dos_2022, mcfadden_sok_2026}, and in wall-clock seconds alongside. Across six ML-NIDS environments, two operating points, both adversarial objectives, and a range of perturbation budgets, we find that break-even arrives after 76 to 1{,}622 adversarial examples in queries, that the sequential formulation the literature adopts by default is not load bearing, and that a policy trained against one detector retains most of its effect against others, lowering $N^{*}$ further. Below the threshold, RL is an expensive way to do what search already does; above it, search is an expensive way to do what a policy has learned.

\noindent\textbf{Contributions.} We formalize the choice between learned and search-based generation as an amortization problem with a measurable break-even point and specify the accounting under which the two are comparable. We measure that break-even across six ML-NIDS environments against gradient, query-based, and random baselines, in detector queries and wall-clock time. We show that the multi-step formulation is unnecessary for most of the threat surface, that the learned policy conditions on its input rather than converging on a fixed perturbation, and that it transfers across detectors and traffic distributions. We release our code, trained policies, and per-configuration query logs at \href{https://anonymous.4open.science/r/rol_artifact-4DD8}{anonymous.4open.science/r/rol\_artifact-4DD8}.

\section{Background \& Motivation}\label{sec:background}
We describe how ML-NIDS are used in practice (\autoref{sec:ml-nids}), summarize adversarial examples in ML-NIDS (\autoref{sec:aml-ml-nids}), and motivate the use of RL to generate adversarial examples against ML-NIDS (\autoref{sec:whyrl}) before defining our research goal.

\subsection{Detecting Intrusions with ML}\label{sec:ml-nids}
Detecting intrusions in a network is a 40 year old problem~\cite{denning_intrusion-detection_1987}. ML-NIDS follow a common pipeline: packets are captured at a vantage point (e.g., a router), aggregated into flows (e.g., SplitCap~\cite{netresec_splitcap_2011}), summarized as feature vectors (e.g., NetFlow~\cite{claise_cisco_2004}), and submitted to a trained ML model whose output determines whether an alert is surfaced to an operator~\cite{apruzzese_sok_2026}. A good NIDS maximizes its true positive rate (TPR), the fraction of malicious flows it flags, while minimizing its false positive rate (FPR), the fraction of benign flows it flags. Minimizing FPR takes priority in practice: the overwhelming majority of alerts triaged are false alarms~\cite{alahmadi_99_2022, tariq_alert_2025}, and the resulting alert fatigue---not missed detections---is the principal reason ML-NIDS are tuned down~\cite{guo_knowml_2026, ede_deepcase_2022}.

\noindent\textbf{ML-NIDS configuration.} We scope our work to supervised ML classifiers over fixed-length feature vectors derived from NetFlow records~\cite{barradas_flowlens_2021, apruzzese_sok_2023, layeghy_benchmarking_2024}. We call a single submission of a feature vector to the detector a \textit{query}: the atomic unit of interaction with an ML-NIDS. At network scale~\cite{edwards_laymans_2025}, both the operator (performing audits) and the adversary aim to minimize the number of model queries, as each costs time, generates traffic, and can risk exposure.

\subsection{Evading ML-NIDS}\label{sec:aml-ml-nids}
\begin{figure}
    \centering
    \includegraphics[width=0.9\linewidth]{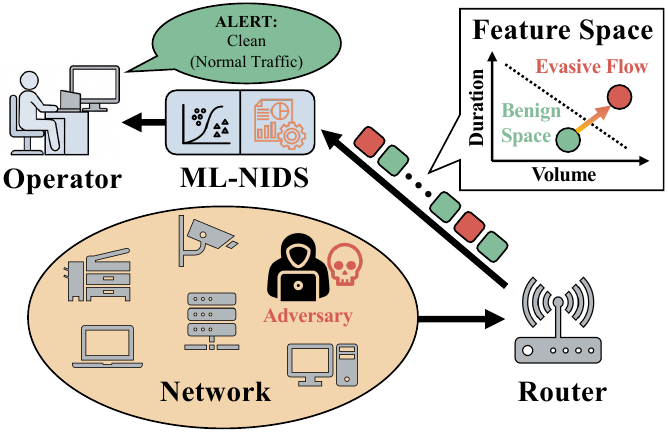}
    \caption{\textbf{Exemplary ML-NIDS evasion attack scenario adapted from}~\cite{apruzzese_sok_2026}. The adversary manipulates traffic in the network that routes to the ML-NIDS where \textit{duration} and \textit{volume} features are indirectly perturbed. The ML-NIDS raises no alert and the operator observes seemingly benign traffic.}
    \label{fig:aml-mlnids}
\end{figure}

Adversarial examples, or inputs perturbed to induce misclassification on ML models, were first characterized on image classifiers~\cite{szegedy_intriguing_2014, biggio_evasion_2013, goodfellow_explaining_2015, papernot_limitations_2016} and have since been demonstrated against essentially every security-relevant learning task~\cite{pierazzi_intriguing_2020, anderson_practical_2021}. They carry a particular importance in ML-NIDS, where the ML model \textit{is} the detector: an adversary who can reliably flip its decision can control what alerts are fired. The community has studied evasion attacks on ML-NIDS almost exclusively as perturbing\footnote{In this domain, perturbations are functionality-preserving packet manipulations, such as introducing inter-packet delays, fragmenting packets, or appending null bytes~\cite{liu_hard-label_2025, nasr_defeating_2021}.} malicious flows to appear benign, driving down TPR. This is half the threat surface. Perturbing benign flows to trigger (many) alerts consumes operator resources, allowing attackers to evade detection via denial-of-service of the ML-NIDS~\cite{ptacek_insertion_1998}. Thus, we treat both instances of adversarial examples in ML-NIDS (malicious$\rightarrow$benign $|$ benign$\rightarrow$malicious) as the same evasion primitive.

Generating an adversarial example is a search problem: find a perturbation, admissible under the adversary's constraints, that changes the model's decision. In the black-box setting this is solved by gradient estimation or direct query-based search, both of which spend a query budget \textit{per input}~\cite{liu_amoeba_2023}. In the image domain this search is expensive because the input is high-dimensional and the perturbation must be located in that space. This explanation does not apply to ML-NIDS. Flow features are low-dimensional and tightly constrained---only a handful of features are realizable by a host-side adversary, and domain constraints bind the rest~\cite{sheatsley_robustness_2021, verkerken_what_2026}---so any \textit{single} instance is comparatively cheap to solve. What is expensive is the population: an adversary does not want one evade once, but every flow in a campaign to evade, and per-input search pays the full cost for each flow. This is where retaining something transferable from each evasive flow could pay dividends.

\noindent\textbf{Scope of this work.} \textit{We generate adversarial examples on ML-NIDS in the feature space}. The adversary observes model outputs and has no access to parameters, architecture, or training data; every model query is counted, including those consumed during RL. We consider two forms of evasion: malicious$\rightarrow$benign ($\downarrow$ TPR) and benign$\rightarrow$malicious ($\uparrow$ FPR). Following ~\cite{apruzzese_evading_2018, nasr_defeating_2021, liu_hard-label_2025}, we restrict the adversary to strictly additive perturbations to the flow \textit{volume} and \textit{duration} features. As illustrated in \autoref{fig:aml-mlnids}, this perturbation crosses the ML-NIDS decision boundary within the feature space.

\subsection{Why RL?}\label{sec:whyrl}

Reinforcement learning (RL) formalizes decision making as a Markov decision process (MDP) $(\mathcal{S}, \mathcal{A}, P, R)$, in which an agent observes a state $s \in \mathcal{S}$, selects an action $a \in \mathcal{A}$, and receives a reward $r$; training seeks a policy $\pi$ maximizing expected return~\cite{sutton_reinforcement_2020}. Applied to adversarial example generation, the state is the input under perturbation together with any feedback the detector returns, the actions are admissible perturbations, and the reward is misclassification~\cite{domico_adversarial_2026, tsingenopoulos_adaptive_2025, tsingenopoulos_autoattacker_2019}. The property that distinguishes RL from search is not how a single input is perturbed but \textit{where the cost is paid}: search spends its budget per input and retains nothing, whereas RL spends a budget once during training and yields a policy applied thereafter at low marginal cost. RL has been used in this way against malware detectors~\cite{anderson_evading_2017, tsingenopoulos_how_2024}, censorship classifiers~\cite{liu_amoeba_2023}, CAPTCHA systems~\cite{tsingenopoulos_captcha_2022}, and ML-NIDS~\cite{wang_crafting_2021, hore_deep_2025, apruzzese_deep_2020}.

\noindent\textbf{The unsettling truth.} Despite this activity, the field has not converged on whether learning is warranted~\cite{mcfadden_sok_2026}. Reported results are difficult to compare: training queries are routinely excluded from the attack's stated cost, multi-step RL formulations are adopted without first testing whether the problem is even sequential at all, and policies are evaluated on the detector and traffic distribution they were trained against~\cite{mcfadden_sok_2026, arp_dos_2022, tsingenopoulos_adaptive_2025, apruzzese_sok_2023}. The consequence is that value proposition of an adversary that learns is unknown, as is the scale at which it beats simply searching again. Adversaries face a real opportunity cost, since budget spent training a policy is budget not spent evading now.

\begin{goal}
    \textbf{Research Goal:} We aim to establish when learning a reusable adversarial policy is preferable to per-input search for generating adversarial examples against ML-NIDS. We hypothesize that its value is amortization, and we measure the cost at which a learned adversary becomes viable at scale, across heterogeneous ML-NIDS and threat models.
\end{goal}

\section{Measuring the Cost of Evasion}\label{sec:cost}
Before introducing any generation method, we define what it costs to run one. This section fixes the unit of cost (\autoref{sec:cost-model}), the space in which perturbations are applied (\autoref{sec:pspace}), and the access an adversary is assumed to have (\autoref{sec:threat}).

\subsection{Cost Model}\label{sec:cost-model}
We model the ML-NIDS as a function $f: \mathcal{X} \to [0,1]$ paired with a decision threshold. Since analysts absorb a bounded number of alerts, operators do not deploy detectors at the threshold maximizing some balanced metric; they deploy at the highest threshold whose alert volume fits the budget they have~\cite{alahmadi_99_2022, tariq_alert_2025}. We therefore calibrate each detector on held out benign data to a target false positive rate $\varphi$, yielding the threshold $\tau_\varphi \in [0,1]$ and the detector
\begin{equation}
    D_\varphi(x) = \mathbb{1}\left[f(x) \ge \tau_\varphi\right].
\end{equation}
Every result in this paper is reported at a stated $\varphi$. Evasion is the drop in true positive rate at fixed $\varphi$, written TPR@$\varphi$, measured on the same malicious inputs before and after perturbation. Alert inflation is the false positive rate induced on benign inputs, measured against the calibrated $\varphi$ rather than against zero. Neither quantity is interpretable without $\varphi$, which is why we define it here and do not restate it.

\noindent\textbf{The unit of cost.} We count \textit{oracle queries} and \textit{wall clock time}. One query is one submission of a feature vector to $D_\varphi$ and one returned decision (i.e., hard-label). Queries are the primary unit because they are spent regardless of how a perturbation is produced, which makes them the only unit under which search and learning are directly comparable. Wall clock time is the secondary unit because it is what the adversary experiences. We count every query and every second a method spends to produce the adversarial examples it reports, including the rollouts consumed during policy training and the surrogate queries consumed when a gradient method cannot differentiate its target (\autoref{sec:search}). We exclude only hyperparameter tuning, which we hold equal across methods (1000 queries per method) and report in \autoref{baseline_optimization}. Every wall clock measurement comes from a single probe job pinned to one core, so times are comparable within a section and only as ratios outside one.

\noindent\textbf{Marginal cost.} Methods differ in how often they succeed, so counting cost per attempt would reward a method that fails cheaply. We instead charge each method for the examples it delivers. For an attack $\mathcal{A}$ run over a set of inputs, let $Q(\mathcal{A})$ and $T(\mathcal{A})$ be the queries and seconds it spent, and $S(\mathcal{A})$ the number of inputs on which the ML-NIDS misclassified. The marginal costs are:
\begin{equation}
    \bar{q}(\mathcal{A}) = \frac{Q(\mathcal{A})}{S(\mathcal{A})},
    \qquad
    \bar{t}(\mathcal{A}) = \frac{T(\mathcal{A})}{S(\mathcal{A})},
\end{equation}
the expected queries and seconds per successful adversarial example. Failures are thus paid for, and a method that succeeds rarely is charged what it wasted.

\noindent\textbf{Amortization.} The two strategies have different cost structures. Per input search begins each input from scratch, so its total cost over $N$ examples is linear. A policy pays $Q_{\text{train}}$ once and is then applied at marginal cost:
\begin{equation}
    C_{\text{search}}(N) = N \cdot \bar{q}_{\text{search}}, \qquad
    C_{\pi}(N) = Q_{\text{train}} + N \cdot \bar{q}_{\pi}.
\end{equation}
Learning is worth its cost only where these cross. We define the \textit{break-even volume} $N^{*}$ as the smallest number of adversarial examples for which the policy is no more expensive than search, and $T^{*}$ as the same quantity in the adversary's time:
\begin{equation}\label{eq:nstar}
    N^{*} = \left\lceil \frac{Q_{\text{train}}}{\bar{q}_{\text{search}} - \bar{q}_{\pi}} \right\rceil,
    \qquad
    T^{*} = \left\lceil \frac{T_{\text{train}}}{\bar{t}_{\text{search}} - \bar{t}_{\pi}} \right\rceil .
\end{equation}
Both are defined only when the policy is cheaper at the margin. Where it is not, the policy never repays its training and we report the environment as such rather than omitting it. Both are properties of the target as much as of the method, so we report them per environment and per baseline rather than aggregating. They also need not agree: a method can be query cheap and time expensive, or the reverse, and where the two disagree we report both and say which resource binds. Since each denominator is a difference of two measured quantities, both are sensitive to how well the search baselines are tuned; we tune them under the budget stated above and report their configurations in \autoref{baseline_optimization}.

\subsection{Perturbation Space and Constraints}\label{sec:pspace}
We perturb the five NetFlow features of \autoref{tab:features}. Given a flow $x \in \mathbb{R}^d$ and a cumulative perturbation $\Delta$ supported on those features, the adversarial input is $x' = x + \Delta$.

\noindent\textbf{Budget $\rho$.} Each perturbable feature carries its own budget: $0 \le \Delta_j \le \rho$ for every $j$, so the feasible set is the box $[0, \rho]^5$. The budget is per feature rather than joint because the five manipulations are independently purchased in the network: appending bytes, injecting packets, and delaying transmission draw on different resources, and a joint cap would let an adversary trade delay for bytes at an exchange rate the network does not offer. The box is non-negative because an adversary can add traffic but cannot retract what it has already sent~\cite{nasr_defeating_2021, apruzzese_evading_2018}. When a perturbation is produced over several actions, the constraint binds on the total rather than on any one of them: the cumulative $\Delta$ is clipped back into $[0,\rho]^5$ before it reaches the detector, so an adversary may approach the budget in one action or in many (\autoref{sec:mdp}).

\noindent\textbf{Distortion.} The budget bounds what is permitted; how much of it a given attack consumes is a separate quantity. We measure consumption as $\lVert \Delta \rVert_2$, which ranges from zero to
\begin{equation}\label{eq:bmax}
    B = \sqrt{5}\,\rho ,
\end{equation}
attained when every feature is displaced to its budget. The ratio $\lVert \Delta \rVert_2 / B$ is the fraction of the budget spent, and it is the term through which the policy is penalized for expensive solutions (\autoref{sec:mdp}). Reporting distortion in $L_2$ while constraining in $L_\infty$ is deliberate: the constraint is what the network permits, the distortion is how conspicuous the result is, and these are not the same question. \autoref{tab:features} lists the perturbable features, their types, and the direction each may move. Features outside this set are held fixed because no host side manipulation modifies them without destroying the flow.

\begin{table}[t]
\centering
\small
\begin{tabular}{llc}
\toprule
\textbf{NetFlow Feature} & \textbf{Type} &
\textbf{Action $\mathcal{A}$} \\
\midrule
\multicolumn{3}{l}{\textit{Perturbable}} \\
\texttt{IN\_BYTES}       & Discrete    &
$\Delta_{\text{in-bytes}}\ \uparrow$ \\
\texttt{OUT\_BYTES}      & Discrete    &
$\Delta_{\text{out-bytes}}\ \uparrow$ \\
\texttt{IN\_PKTS}        & Discrete    &
$\Delta_{\text{in-pkts}}\ \uparrow$ \\
\texttt{OUT\_PKTS}       & Discrete    &
$\Delta_{\text{out-pkts}}\ \uparrow$ \\
\texttt{FLOW\_DURATION\_MILLISECONDS}$^{\dagger}$ & Continuous &
$\Delta_{\text{delay}}\ \uparrow$ \\
\midrule
\multicolumn{3}{l}{\textit{Fixed}} \\
\texttt{L4\_SRC\_PORT}   & Categorical & -- \\
\texttt{L4\_DST\_PORT}   & Categorical & -- \\
\texttt{PROTOCOL}        & Categorical & -- \\
\texttt{L7\_PROTO}       & Categorical & -- \\
\texttt{TCP\_FLAGS}      & Categorical & -- \\
\bottomrule
\end{tabular}%
\caption{NetFlow feature space~\cite{sarhan_netflow_2021}. All ten features are
observed by the agent; the action is a $5$-vector, and the cumulative
displacement of every perturbable feature is non-negative ($\uparrow$).
$^{\dagger}$Duration is never set directly: the fifth action component raises a
latent per-packet delay, and duration is re-derived as delay $\times$ (total
packets $-1$). Byte counters are additionally clipped so bytes-per-packet stays
inside the benign $1$st--$99$th percentile band.}
\label{tab:features}
\end{table}

\subsection{Threat Model}\label{sec:threat}

The adversary submits a flow and observes whether an alert was raised. It has no access to scores, parameters, architecture, gradients, or training data, and every submission is charged as a query (\autoref{sec:cost-model}). This is weaker than the access assumed by most of the evasion literature~\cite{sheatsley_robustness_2021, diallo_sabre_2024}, and it reflects what an attacker observes: consequences, not confidences~\cite{apruzzese_real_2022}. We generate perturbations in the feature space and do not synthesize the traffic that would produce them, so what we measure is the cost of \textit{identifying} a perturbation that defeats the detector. Realizing one on a controlled host is a separate and open problem~\cite{verkerken_what_2026}, and one that requires knowing the target first.

\noindent\textbf{The detector queried need not be the detector deployed.} Detection in this domain is homogeneous: operators draw from a small set of classifier families over a standardized NetFlow feature set~\cite{sarhan_netflow_2021, apruzzese_sok_2023}. An adversary can therefore learn a perturbation strategy against a detector of its own and apply it elsewhere. We evaluate both conditions rather than assume one: under \textit{direct} access the adversary queries the detector it attacks; under \textit{transfer} access it queries one detector, or one trained on different traffic, and applies the result without further training. Transfer is what makes $N^{*}$ a property of an adversarial campaign rather than of a single detector.
\section{From Search to Policy}\label{sec:search-to-policy}

We now describe the two attacks that \autoref{sec:cost} makes comparable. We formalize per input search first (\autoref{sec:search}), because it is what a policy must beat, then give the policy formulation (\autoref{sec:mdp}) and show what the horizon does and does not buy (\autoref{sec:collapse}).

\subsection{Per Input Search}\label{sec:search}
Fix a detector $D_\varphi$ and a flow $x$ with label $y$. The adversary seeks a perturbation $\delta$ satisfying the budget from \autoref{sec:pspace} that flips the detector's decision:
\begin{equation}\label{eq:search-obj}
    \text{Find } \delta, \ \lVert \delta \rVert \le \rho, \ \text{ s.t. } \ D_\varphi(x + \delta) \neq y
\end{equation}
Prior work solves \autoref{eq:search-obj} by search: a candidate $\delta$ is proposed, submitted to the detector, and revised according to the response, until the decision flips or a query cap $B$ is reached. We evaluate three search procedures spanning the access assumptions the literature makes.

\noindent\textbf{PGD.}~\cite{madry_towards_2017} Projected gradient descent takes repeated steps along the gradient of the detector's loss, projecting back into the budget after each~\cite{madry_towards_2019}. It is the standard optimizer in the constrained domain literature and the one against which ML-NIDS robustness is usually reported~\cite{sheatsley_robustness_2021, sheatsley_adversarial_2022, jin_robustifying_2025}. It therefore requires direct access to a differentiable detector.

\noindent\textbf{HSJA.}~\cite{chen_hopskipjumpattack_2020} HopSkipJumpAttack estimates the gradient at the decision boundary from the labels returned by the detector, then steps along the estimate. It assumes only that the detector returns a decision, which makes it the strongest baseline available against detectors PGD cannot differentiate, and it is the query based method against which black box evasion of traffic classifiers is typically measured~\cite{liu_hard-label_2025}.

\noindent\textbf{Random search.} Perturbations are drawn uniformly from the budget and submitted until one succeeds or $B$ is exhausted. It assumes nothing about the detector and is included because a policy that does not beat it has learned nothing~\cite{arp_dos_2022}.

Neither of our detector models, random forest and XGBoost, is differentiable, so PGD cannot be run against them directly. Following standard practice we generate perturbations on a differentiable surrogate and transfer them~\cite{papernot_limitations_2016, apruzzese_evading_2018}, which charges PGD the surrogate's queries and costs it whatever the transfer loses. We do not present this as a fair fight: it is the disadvantage deployment reality imposes, since ensemble detectors are the ones operators use~\cite{apruzzese_sok_2023}, and we therefore report the three methods separately throughout \autoref{sec:rq2} rather than averaging gradient based and query based baselines together. What they share is what the paper turns on. Each consumes queries in proportion to the difficulty of the individual input, and every one is charged. Each is also initialized at $\delta = 0$ for every input and retains nothing across inputs: whatever the procedure discovers about the geometry of $D_\varphi$ while solving $x$ is discarded before it begins $x'$. Search is the deployment loop of \autoref{fig:overview} run from scratch for every flow, so its cost over $N$ examples is $N \cdot \bar{q}_{\text{search}}$.

\subsection{Policy Formulation}\label{sec:mdp}

\begin{figure}[t]
    \centering
    \includegraphics[width=\columnwidth]{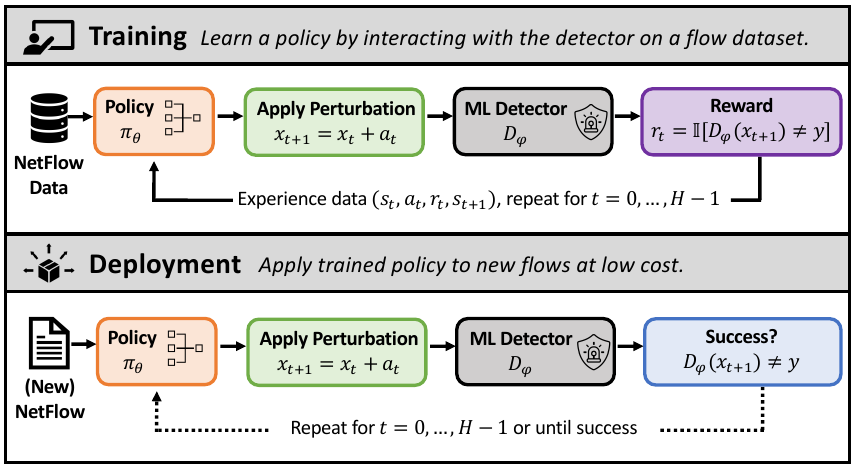}
    \caption{Adversarial example generation with a learned policy. During \textit{training}, the policy interacts with the detector over a flow dataset, consuming one query per step; this cost is $Q_{\text{train}}$, the intercept of \autoref{eq:nstar}. During \textit{deployment}, the trained policy is applied to new flows, spending queries only to confirm success. Search has no training phase and repeats the deployment loop from scratch for every flow.}
    \label{fig:overview}
\end{figure}

The alternative is to solve \autoref{eq:search-obj} once, in the form of a function that maps any flow to a perturbation. \autoref{fig:overview} shows the two phases this creates, and where each spends queries. We formulate the problem as a finite horizon MDP $(\mathcal{S}, \mathcal{A}, P, R, H)$ and keep the formalism minimal, since the contribution of this paper is to establish the role of learning.

\noindent\textbf{States.} A state $s_t \in \mathcal{S}$ comprises the original flow vector, the 5-dimensional perturbation vector accumulated (\autoref{tab:features}), and the remaining perturbation budget available. This is the information available to the adversary, and this way a trained policy need only the flow vector to generate perturbations.

\noindent\textbf{Actions.} An action $a_t \in \mathcal{A} \subseteq \mathbb{R}^5$ is a signed displacement of the mutable features, bounded per feature by the step scale $\lVert a_t \rVert_\infty \le \rho$. Transitions are deterministic and act on the cumulative perturbation rather than on the flow:
\begin{equation}\label{eq:transition}
    \Delta_t = \mathrm{clip}\!\left(\Delta_{t-1} + a_t,\ 0,\ \rho\right), \qquad x_{t+1} = x_0 + \Delta_t ,
\end{equation}
with $\Delta_0 = 0$. Individual actions may therefore be negative, but the clip enforces $\Delta_t \in [0,\rho]^5$ at every step, so what reaches the detector is never a retraction of traffic the adversary has already sent (\autoref{sec:pspace}). A negative component walks back an earlier overshoot within an episode; it does not purchase a perturbation the network could not produce. Because the bound is per feature and equal to the budget, a single action reaches anything the budget permits, and \autoref{sec:collapse} states what this does to the horizon.

\noindent\textbf{Reward.} The adversary observes only the decision, so the reward is the objective itself, discounted by the share of the budget the perturbation consumed (\autoref{sec:pspace}):
\begin{equation}\label{eq:reward}
    r_t = \mathbb{1}\!\left[D_\varphi(x_{t+1}) \neq y\right]\left(1 - \frac{\lVert \Delta_t \rVert_2}{B}\right).
\end{equation}
The indicator is taken on malicious inputs for evasion and on benign ones for inflation. This is the most conservative setting available: a flip that spends the entire budget scores zero, exactly as a failure does, so nothing we report is a success bought by spending everything. We do not shape the reward with detector confidence, which is unavailable under our threat model, nor with distance to the boundary, which would presume knowledge the adversary does not have.

An episode runs for at most $H$ steps and terminates early on success. The same loop runs in both phases of \autoref{fig:overview}; what differs is that training consumes the transitions $(s_t, a_t, r_t, s_{t+1})$ to update $\pi_\theta$, while deployment leaves the policy fixed. Every environment interaction issues exactly one detector query, so cumulative training queries equal cumulative environment steps and form the intercept $Q_{\text{train}}$ of \autoref{eq:nstar}. Algorithms, architectures, training length, and the tuning budget, which we hold equal to that given the search baselines above, are stated in \autoref{baseline_optimization}.

\subsection{What Multi-Step RL Buys}\label{sec:collapse}
Multi step formulations are standard in this literature and are typically adopted without argument. Under \autoref{eq:transition} the horizon buys less than it appears to, and we state the condition precisely because our own results depend on it. Since each action is bounded by $\rho$ and the clip holds $\Delta_t$ inside $[0,\rho]^5$ at every step, the reachable set is
\begin{equation}\label{eq:reachable}
    \mathcal{R}_H(x) = \left\{ x + \Delta \ : \ \Delta \in [0,\rho]^5 \right\}
\end{equation}
for every $H$: one action already reaches everything the budget permits, and additional actions expand nothing. Any advantage measured for $H > 1$ is therefore not a horizon effect, and prior comparisons that cap the per step displacement below the budget are measuring that cap rather than a sequence. What $H>1$ genuinely provides is the $H-1$ intermediate hard labels the adversary observes along the way. The horizon is an information channel, not a trajectory. This is why we formulate the policy as an MDP and then measure whether the MDP was necessary (\autoref{sec:horizon}). Where that channel carries little information, the problem reduces to a contextual bandit over $\mathcal{X}$, and the learned object is a reusable function rather than a learned sequence.
\section{Evaluation}\label{sec:evaluation}
We evaluate whether learning is worth its cost against ML-NIDS. We describe our setup (\autoref{setup}) and then ask:

\begin{rqlist}[series=rquestion]
\item Does learning improve the adversary's ability to generate adversarial examples on ML-NIDS?\label{rq:learning} (\autoref{sec:rq1})
\item Does learning amortize the cost of attacking an ML-NIDS, and at what scale?\label{rq:amortization} (\autoref{sec:rq2})
\item What does the policy learn, and where else does it apply?\label{rq:transfer} (\autoref{sec:rq3})
\end{rqlist}

\subsection{Experimental Setup}\label{setup}

\noindent\textbf{Datasets.} We use three NetFlow intrusion detection datasets curated to a standardized feature set~\cite{sarhan_towards_2022}, together representing 21 network attacks. \texttt{UNSW-NB15}~\cite{moustafa_unsw-nb15_2015} generates a hybrid of real background activity and synthetic attacks with the IXIA PerfectStorm tool, spanning nine families including Fuzzers, Backdoors, and Exploits. \texttt{ToN-IoT}~\cite{alsaedi_ton_iot_2020} collects telemetry from IIoT sensors, operating systems, and network traffic across virtual machines and physical devices, under attacks such as Ransomware, XSS, and Injection. \texttt{CSE-CIC-IDS2018}~\cite{sharafaldin_toward_2018} drives abstract user profiles to produce realistic cloud-centric background traffic over ten days, with Injection, Botnet, Brute Force, and Infiltration scenarios.All three carry labeling and duplication defects~\cite{engelen_troubleshooting_2021, flood_bad_2024}: we deduplicate on the ten model features, retain records whose labels conflict after deduplication, and report the resulting TPR ceiling in \autoref{data_prep}. None carries a usable timestamp, so we split data using stratified sampling.

\noindent\textbf{Models and calibration.} We evaluate the two ensemble families popular in flow-based ML-NIDS, Random Forest (RF) and XGBoost (XGB)~\cite{apruzzese_sok_2023}, across the three datasets, giving six environments. Neither admits a gradient, which is a reality rather than a limitation of our setup. Hyperparameters are selected by TPR at the target FPR on validation (\autoref{detector_sweep}), and each detector is calibrated on a benign-only holdout to $\varphi = 1\%$ and $\varphi = 0.1\%$. \autoref{tab:detectors} reports the pre-attack TPR@$\varphi$ that bounds what any attack can suppress, and two of its properties govern how the rest of this section reads. First, detector model is nearly irrelevant: RF and XGB agree to within 0.05 points, so variation across environments is driven by traffic, not architecture. Second, the operating point is decisive. On UNSW-NB15, tightening the alert budget by a factor of ten costs up to 0.4 TPR, from 0.917 to 0.514: nearly half the malicious traffic goes undetected before any adversary intervenes. A metric averaged over thresholds would call this detector strong, but at the threshold an analyst can absorb, it is not.

\begin{table}[t]
\centering
\setlength{\tabcolsep}{4pt}
\begin{tabular}{llrr}
\toprule
\textbf{Dataset} & \textbf{Detector} & \textbf{TPR@1\%} & \textbf{TPR@0.1\%} \\
\midrule
\multirow{2}{*}{ToN-IoT}
 & RF  & $0.9999$ & $0.9987^{\dagger}$ \\
 & XGB & $0.9999$ & $0.9984^{\dagger}$ \\
\midrule
\multirow{2}{*}{CSE-CIC-IDS2018}
 & RF  & $0.8728$ & $0.8631$ \\
 & XGB & $0.8748$ & $0.8633$ \\
\midrule
\multirow{2}{*}{UNSW-NB15}
 & RF  & $0.9172$ & $0.5136$ \\
 & XGB & $0.9182$ & $0.5667$ \\
\bottomrule
\end{tabular}
\caption{Detector operating points. Each detector is calibrated to target false positive rates of 1\% and 0.1\%. We report the pre-attack detection rate (TPR), which upper-bounds what any attack can suppress. $^{\dagger}$ToN-IoT's benign holdout supports a $0.1\%$ threshold only marginally; see Appendix \autoref{detector_sweep}. We report TPR rather than F1: a detector is deployed at a fixed alert budget, and averaging over thresholds describes a configuration no operator runs~\cite{apruzzese_sok_2023}.}
\label{tab:detectors}
\end{table}

\noindent\textbf{Algorithms and partitioning.} We train policies with PPO~\cite{schulman_proximal_2017}, SAC~\cite{haarnoja_soft_2018}, and A2C~\cite{mnih_asynchronous_2016} under stable-baselines3~\cite{raffin_stable-baselines3_2021}, spanning on-policy and off-policy training. The loops differ, but the deployed object is the policy $\pi_\theta$, and every environment step issues exactly one detector query regardless of algorithm. Each dataset is split four ways: $\mathcal{D}_{\text{target}}$ trains the detector, $\mathcal{D}_{\text{cal}}$ (benign only) calibrates its threshold, $\mathcal{D}_{\text{train}}$ trains the policy against the calibrated detector, and $\mathcal{D}_{\text{test}}$ measures both attack effectiveness and attack cost for fair comparison~\cite{arp_dos_2022}.

\noindent\textbf{Metrics and baselines.} We measure evasion as post-attack TPR@$\varphi$ and alert inflation as the FPR induced on benign flows, both against the calibrated operating point (\autoref{sec:cost-model}). We do not report ASR or accuracy: a detector is deployed at a fixed alert budget, and metrics that average over thresholds describe a configuration no operator runs~\cite{apruzzese_sok_2023}. Cost is reported as $\bar{q}$, $\bar{t}$, $N^*$, and $T^*$. We benchmark against the three search procedures of \autoref{sec:search}: PGD, which reaches RF and XGBoost only through an MLP surrogate; HSJA, which queries any target directly under hard-label access; and random search at a matched budget. All methods are subject to the same constraints (\autoref{sec:pspace}) and a query cap of $Q_{\max} = 1000$ per input. Search algorithm hyperparameters are tuned for queries-to-success under the query budget.

\subsection{Attacking ML-NIDS with a Learned Policy}\label{sec:rq1}
We first ask \ref{rq:learning}: \textit{does learning improve the adversary's ability to generate adversarial examples on an ML-NIDS?} An adversary considering a learned policy needs two things settled before cost is worth discussing. It needs to know that the policy induces failure in a detector at the threshold that detector is actually deployed at, and it needs to know which tuning parameters make it successful. We answer the first in \autoref{sec:effectiveness}, sweeping three learning algorithms across all six environments and both operating points, and the second in \autoref{sec:horizon}, sweeping the perturbation budget and the episode horizon. Throughout, the policy acts once per flow unless the horizon is the control variable.

\subsubsection{Effectiveness}\label{sec:effectiveness}

\begin{figure}[t]
    \centering
    \includegraphics[width=\linewidth]{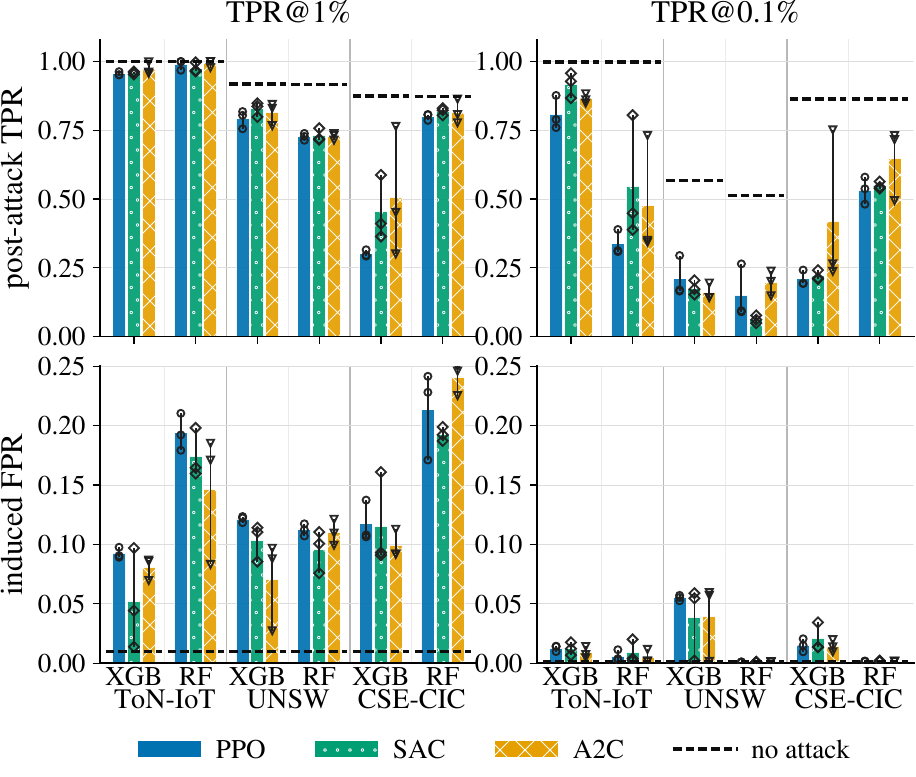}
    \caption{Post-attack detection across all six environments, at both operating points, for three learning algorithms. Top row is retained TPR@$\varphi$ on malicious flows; bottom row is the FPR induced on benign flows. Dashed rules are the detector before attack. Markers are the three seeds, whiskers their min and max. Note the right column: every environment is easier to evade at $\varphi = 0.1\%$ than at $\varphi = 1\%$.}
    \label{fig:effectiveness}
\end{figure}

We train policies with PPO, SAC, and A2C in each of the six environments, at both operating points and under both objectives, for $8\times10^4$ environment steps at $\rho = 0.08$, over three seeds and evaluated on $10^4$ held out episodes (\autoref{baseline_optimization}). \autoref{fig:effectiveness} reports the result.

The attack works, but unevenly. At $\varphi = 1\%$ it suppresses CSE-CIC XGBoost from $0.87$ to $0.30$ and UNSW RandomForest from $0.92$ to $0.73$, while ToN-IoT resists almost entirely at $0.95$ and above. Inflation is uniform where evasion is not: every environment is driven to between $0.09$ and $0.24$ induced FPR, nine to twenty four times the alert volume the detector was calibrated to produce. The choice of algorithm is immaterial, with PPO, SAC, and A2C inside each other's seed spread everywhere but CSE-CIC XGBoost.

The security finding is in the right column. Tightening the alert budget by a factor of ten does not make a detector harder to evade, it makes it easier. ToN-IoT RandomForest, which barely moves at $\varphi = 1\%$, falls from $0.999$ to $0.33$ at $\varphi = 0.1\%$; UNSW XGBoost falls from $0.567$ to $0.21$. The mechanism is direct: a tighter alert budget raises the threshold, a higher threshold sits closer to the flows the detector was already alerting on, and less displacement is needed to carry a malicious flow across it. The environments that resist attack are the ones with room for false positives, which is exactly what an operator gives up to control false alarms.

This is dangerous for a deployed ML-NIDS. Operators do not choose their threshold freely: analyst capacity chooses it for them, and the pressure runs toward fewer alerts (\autoref{sec:ml-nids}). That pressure produces the configuration an adversary prefers, and the two objectives then compound. An adversary that inflates alert volume twenty fold forces exactly the response that makes evasion cheaper, and a defender who tightens against the flood has moved to the threshold at which the flood's author evades most easily. Denial of service against a NIDS has been an evasion strategy since Ptacek and Newsham~\cite{ptacek_insertion_1998}; what is new is that the two are not alternatives but a sequence a single learned policy can exploit.

\resultline{A zero-shot learned policy evades detection in five of six environments and multiplies alert volume nine to twenty four fold. Evasion is consistently more severe at the tighter operating point, so the threshold an alert budget forces on an operator is the one an adversary can defeat most easily.}

\subsubsection{Budget and Horizon}\label{sec:horizon}
\begin{figure}[t]
    \centering
    \includegraphics[width=\linewidth]{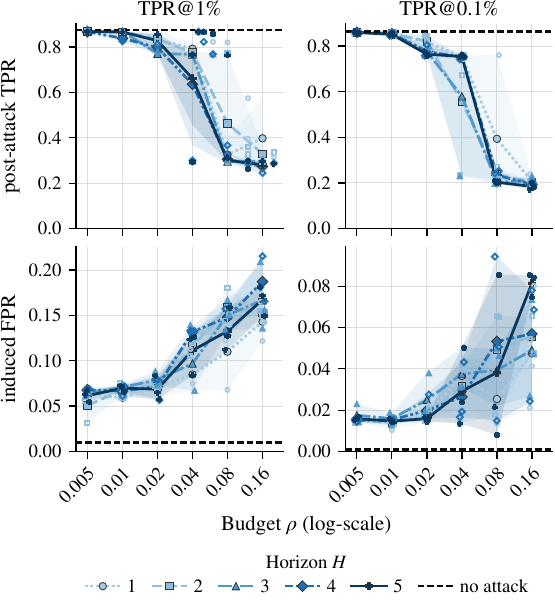}
    \caption{Post-attack TPR@$\varphi$ (top) and induced FPR (bottom) against $\rho$ on a log axis, one line per horizon $H$, bands over three seeds. Detection is untouched below $\rho = 0.02$ and collapses by $0.08$. The five horizons are indistinguishable for evasion at (almost) every budget.}
    \label{fig:horizon_budget}
\end{figure}

Effectiveness is not a property of the policy alone. It depends on how far the adversary is permitted to move, $\rho$, which is set by what the network tolerates and is the defender's lever, and on how many attempts it is permitted to make, $H$, which is chosen by the adversary and is the assumption the RL literature makes without testing~\cite{mcfadden_sok_2026}. We sweep $H \in \{1,\ldots,5\}$ against six budgets from $\rho = 0.005$ to $0.16$ at log-scale, at both operating points and under both objectives. Because every action may reach $\rho$ on every feature, additional steps buy \textit{retries}, not \textit{reach} (\autoref{sec:collapse}), so what the horizon is tested for here is whether intermediate feedback is worth anything. The sweep is 360 runs and is necessarily narrower than \autoref{sec:effectiveness}: it covers PPO on CSE-CIC XGBoost only, and we make no claim that the horizon behaves identically elsewhere.

The budget behaves as a cliff rather than a slope. Below $\rho = 0.02$ nothing happens: post attack TPR sits on the pre attack rule and the detector is untouched. Between $0.02$ and $0.08$ it collapses from $0.85$ to $0.30$, and past $0.08$ it flattens. Over a factor of four in budget, the adversary goes from no useful attack to a complete one.

The horizon buys almost nothing. For evasion the five lines are indistinguishable at every budget, and $H = 1$ is marginally better at $\rho = 0.08$. For inflation a separation appears only at the largest budgets, where $H = 4$ reaches $0.19$ induced FPR against $H = 1$'s $0.15$, and it is absent when the budget is tight. This is what retries rather than reach look like. Evasion needs one displacement in the cheapest available direction and a single action finds it, while inducing a false positive means landing a benign flow beyond a threshold calibrated to fire on one percent of traffic, a smaller target that repeated attempts help with only once it is within reach.

What this says is that the \textit{sequential} formulation is unnecessary, not that learning is. At $H = 1$ the environment reduces to a contextual bandit: the policy still learns a mapping from flows to perturbations from detector feedback, and \autoref{sec:universal} shows that mapping is not a constant. What disappears is the trajectory and the credit assignment that comes with it~\cite{sutton_reinforcement_2020}. For an adversary this is a strict improvement, since it removes $H-1$ detector interactions per flow at deployment, each a cost in queries and time. For the RL for security community it is a correction: multi step formulations are the default in this domain~\cite{hore_deep_2025, liu_amoeba_2023, apruzzese_deep_2020}, and on the axis we can test they are not worth their cost.

For a defender, the cliff is the actionable finding. Adversarial robustness in ML-NIDS is usually treated as a property of the model~\cite{jin_robustifying_2025}; \autoref{fig:horizon_budget} suggests it is at least as much a property of how much displacement the network tolerates before a flow stops serving its purpose. An operator that holds an adversary below $\rho \approx 0.02$ has a robust detector and one that cannot hold it below $0.08$ has an extremely vulnerable one, with very little ground in between.

\resultline{Attack success is governed by the perturbation budget, which produces a cliff between $\rho = 0.02$ and $0.08$, and not by the horizon, which is indistinguishable across $H$ for evasion and separates only for inflation at the largest budgets.}

\begin{goal}
\textbf{Answer to \ref{rq:learning}:} Learning gives the adversary a zero shot attack on heterogeneous ML-NIDS, evading detection in five of six environments and multiplying alert volume up to twenty four fold, without gradients and against detectors no gradient method reaches directly. It works best at the tighter operating point, so the configuration alert fatigue forces on an operator is the one an adversary prefers. It does not require a sequence.
\end{goal}

\subsection{Amortizing the Cost of Attack}\label{sec:rq2}
\autoref{sec:rq1} showed that a learned policy breaks ML-NIDS detectors at the thresholds they are deployed at. It did not motivate that anyone should train one. An adversary holding a search algorithm that already works has a fair objection, and answering it requires pricing both options in the same units. Here we answer \ref{rq:amortization}: \textit{does learning amortize the cost of attacking an ML-NIDS, and at what scale?} We proceed in two steps. \autoref{sec:marginal} measures what one successful adversarial example costs each method once it is running, which fixes the slope of the comparison under our cost units defined in \autoref{sec:cost-model}. \autoref{sec:amortization} adds the training cost the policy pays before it runs, which fixes the intercept, and reports where the two cross. Queries are counted on the compute cluster and pool freely across nodes. Seconds do not: every wall clock measurement in this section comes from a single probe job pinned to a single core (\autoref{timing_env}), so times are comparable within this section and only as ratios against anything outside it.

\subsubsection{Marginal Cost}\label{sec:marginal}
\begin{figure}
    \centering
    \includegraphics[width=\linewidth]{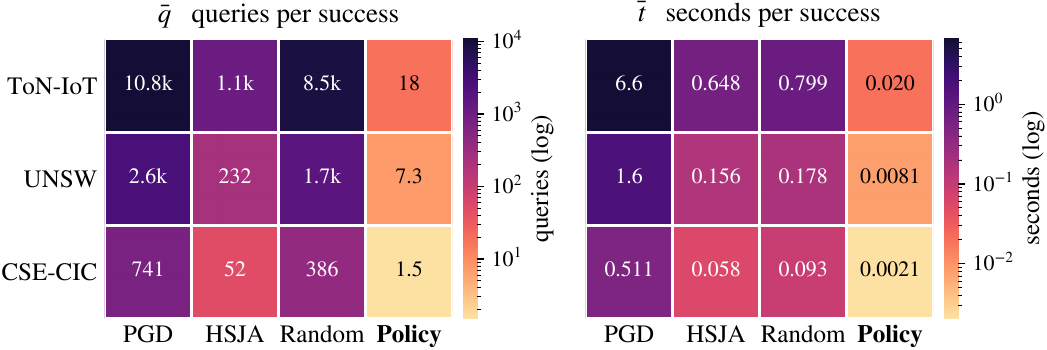}
    \caption{Cost of one successful adversarial example, in target queries (left) and wall-clock seconds (right). The policy is $61\times$ cheaper than HSJA on ToN-IoT and reaches $1.5$ queries per success on CSE-CIC, roughly one confirmation query at a two-in-three success rate.}
    \label{fig:marginal_cost}
\end{figure}

Attack papers report how often an attack succeeds, but an adversary needs to know what a success costs, and a method that succeeds nineteen times in twenty at a hundred queries each is a different proposition from one that succeeds once in twenty at ten. We therefore divide each method's total expenditure by the successful adversarial examples it produced, so a method that wastes its budget on flows it never flips pays for that budget in full. We run PGD, HSJA, and random search against 500 flows per environment at $Q_{\max} = 1000$, alongside the trained policy, on XGBoost victims at $\varphi = 1\%$ and $\rho = 0.08$, charging PGD for every query on the surrogate and on the victim alike.

\begin{figure*}[t]
    \centering
    \includegraphics[width=\linewidth]{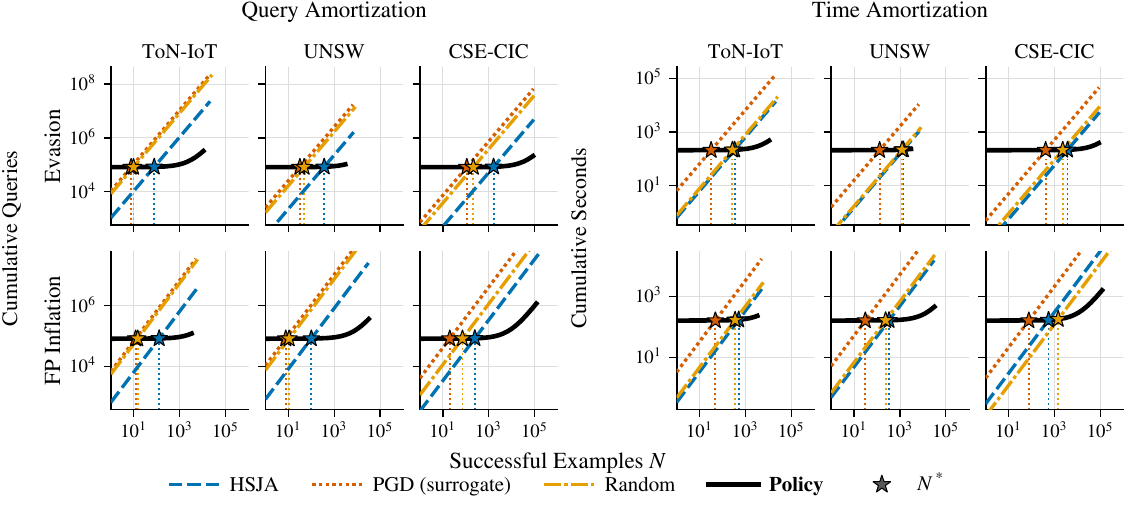}
    \caption{Where paying for training up front becomes the cheaper decision. Cumulative cost against successful examples produced, in queries (left block) and seconds (right block), for evasion (top) and alert inflation (bottom). Search lines start at the origin, the policy starts at its training cost (81{,}920 queries, 209.2s) and rises at the marginal rates of \autoref{fig:marginal_cost}. $N^{*}$ marks each crossing; a line that ends has hit the production ceiling, beyond which the environment cannot supply more successes.}
    \label{fig:amortization}
\end{figure*}

The policy is between one and three orders of magnitude cheaper than every search baseline in every cell of \autoref{fig:marginal_cost}. Against the strongest baseline, HSJA, it is $61\times$ cheaper in queries on ToN-IoT ($18$ against $1{,}100$), $32\times$ on UNSW, and $35\times$ on CSE-CIC. The ordering of environments is the one \autoref{sec:rq1} showed, with ToN-IoT the most expensive target for every method and CSE-CIC the cheapest: cost and difficulty are the same axis viewed twice. On CSE-CIC the policy reaches $1.5$ queries per success, close to the floor of one confirmation query at a two-in-three success rate.

The surprising row is PGD. It is the most expensive method in both currencies and in every cell, and random sampling from the perturbation budget beats it on queries everywhere: $8.5$k against $10.8$k on ToN-IoT, $1.7$k against $2.6$k on UNSW, $386$ against $741$ on CSE-CIC. Part of this is an artifact of reaching a non differentiable victim, since roughly 17\% of its queries land on the detector and the rest on the surrogate, so its exposure to a rate limited target is smaller than its query cell implies. The seconds panel is the check on that excuse, and PGD remains last: gradient descent through a surrogate is expensive in compute whether or not the queries reach the victim. The standard optimizer in this literature is, per successful example, the worst of the four we tested.

For a defender this reframes what query volume can be used for. A detector that logs repeated probes has a plausible signal against HSJA, which needs $1{,}100$ interactions to produce one evasive flow on ToN-IoT and would be conspicuous doing it~\cite{stateful_carlini}. It has no such signal against a policy that needs eighteen, and effectively none against one that needs $1.5$. The detection strategy that works against query based attacks is the one a learned adversary makes irrelevant, because it has moved its interaction volume off the victim and into a training phase the victim never observes.

\resultline{One successful adversarial example costs the policy $1.5$ to $18$ target queries, against $52$ to $1{,}100$ for the strongest search baseline. PGD is the most expensive method in both currencies, and random sampling beats it on queries in every environment.}

\subsubsection{When Training Pays for Itself}\label{sec:amortization}

The marginal costs above are slopes, and a slope alone decides nothing. The policy owes 81{,}920 queries and 209.2 seconds before it produces its first adversarial example, and search owes nothing, so whether learning is the right decision depends on how many examples the adversary intends to produce. We accumulate cost against successful examples for each method, in both currencies and under both objectives, charging training in full and once against the row's own timing probe (\autoref{fig:amortization}).

Every crossing is early. In queries, the policy overtakes HSJA after 76 successful examples on ToN-IoT, 365 on UNSW, and 1{,}622 on CSE-CIC; against PGD and random search it overtakes in tens of flows. The ordering is the reverse of marginal cost, for a reason worth stating: break-even is latest where the attack is cheapest, because a cheap environment gives search less to lose. CSE-CIC, the environment a learned adversary attacks most easily, is where learning takes longest to pay for itself. None of this is a meaningful barrier at the scale these detectors operate at. Public benchmarks contain tens of millions of records and a monitored link produces them continuously (\autoref{sec:ml-nids}), so an adversary crosses even the latest threshold inside a few seconds of traffic.

The two currencies do not agree, and the disagreement is against us. Break-even in seconds is four to five times larger than in queries: 333 examples on ToN-IoT rather than 76, 1{,}414 on UNSW, 3{,}743 on CSE-CIC. The cause is not a defect in either measurement. Training issues its queries at roughly 390 per second against HSJA's 1{,}700, so the policy's training phase is slower per interaction than the search it replaces. An adversary counting its exposure to a detector therefore reaches break-even sooner than one counting hours on its own machine, and the honest reading is that queries are the currency in which our result is strongest. We report both because an adversary under time pressure and an adversary under rate limiting are different adversaries, and only one is well described by a query count~\cite{merkle_economics_2024}.

The security consequence is not that a learned attack exists, which prior work established~\cite{hore_deep_2025, liu_amoeba_2023}, but that its economics are settled far below the scale of any campaign worth mounting. The objection that has kept learned adversaries out of practice is that training is a large fixed cost paid before any return~\cite{apruzzese_real_2022}; at break-even points in the hundreds it is amortized within the first minute of a campaign, after which the adversary holds a reusable artifact that produces adversarial flows at a cost the detector cannot distinguish from ordinary traffic. A defender cannot raise these thresholds by making its detector harder to query, because the queries are not spent on the detector. It can raise them only by narrowing the perturbation budget, which \autoref{sec:horizon} shows collapses the attack below $\rho = 0.02$, or by shortening the useful life of a trained policy, which \autoref{sec:rq3} tests.

\resultline{The policy amortizes its training cost after 76 to 1{,}622 successful examples in queries, and 333 to 3{,}743 in wall clock seconds, against benchmarks with tens of millions of flows. Break-even arrives latest where the attack is easiest.}

\begin{goal}
\textbf{Answer to \ref{rq:amortization}:} Yes, at a scale that removes the objection to the approach. Training is recovered after 76 to 1{,}622 successful examples in queries and 333 to 3{,}743 in seconds, every threshold orders of magnitude below the volume of a single intrusion detection benchmark, and past it the policy is cheaper without limit.
\end{goal}

\subsection{What the Policy Learns}\label{sec:rq3}
\autoref{sec:rq2} showed that training pays for itself within hundreds of adversarial examples. That result is only meaningful if the policy acquired something worth paying for. Two failure modes would hollow it out, and they pull in opposite directions. If the policy has collapsed onto a single fixed perturbation, then it has learned a constant, and an adversary could have found that constant far more cheaply than by training a network. If instead it has memorized one detector, it is worthless the moment the target changes, and the amortization of \autoref{sec:amortization} applies to a campaign of one. Here we answer \ref{rq:transfer}: \textit{what does the policy learn, and where else does it apply?} \autoref{sec:universal} examines what the policy does to a flow and shows it is a function of that flow, not a constant. \autoref{sec:transfer} moves trained policies to detectors and traffic they never saw. The two must be read together: a policy that had collapsed to a constant would transfer trivially, so a transfer result alone proves nothing about what was learned.

\subsubsection{The Policy Is Not a Universal Perturbation}\label{sec:universal}

\begin{figure}[ht]
    \centering
    \includegraphics[width=0.87\linewidth]{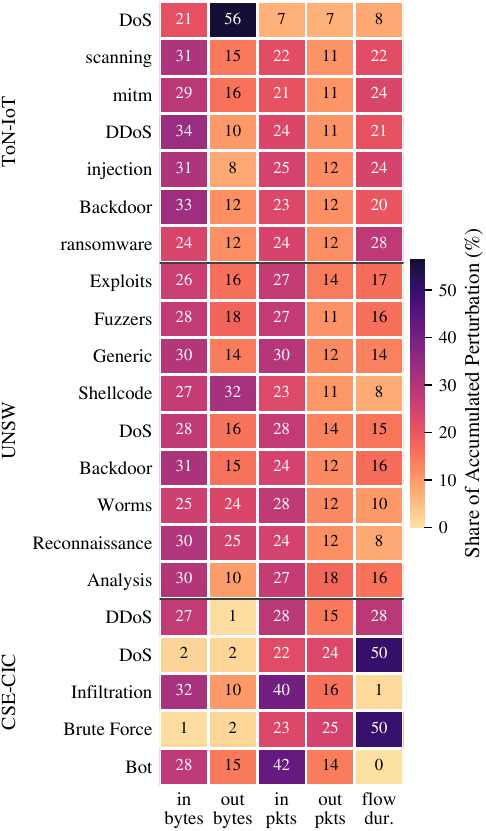}
    \caption{One policy, twenty one attack classes, no universal perturbation. Each column is the share of an attack class's perturbation budget the policy spent on each of the five mutable NetFlow features; columns sum to 100\%, so this is the shape of the perturbation and not its size. Classes are ordered by attack success within each dataset. Labels are the source datasets' own and are not synchronized across each dataset.}
    \label{fig:perturbations}
\end{figure}

A single perturbation that evades a detector on most inputs is a known phenomenon in other domains~\cite{moosavi-dezfooli_universal_2017}, and it would be devastating here: such a perturbation amortizes perfectly and costs almost nothing to find, so a policy converging on one would make the training cost of \autoref{sec:amortization} a purchase with nothing behind it. We therefore examine not whether the policy succeeds, which \autoref{sec:rq1} settles, but what it produces. We log every perturbation during evaluation, decompose it across the five mutable features, and group episodes by the attack class the source dataset assigns to the flow, normalizing each class by its total spend so what we report is the \textit{shape} of the perturbation rather than its size. One policy is trained per environment, so every column within a dataset block of \autoref{fig:perturbations} is the same network responding to different traffic.

The allocations are not the same. In CSE-CIC, DoS receives half its budget on flow duration and only two percent on incoming bytes, while Bot receives forty two percent on incoming packets and nothing at all on duration; Brute Force follows DoS and Infiltration follows Bot. These are four columns produced by the same network, and two are close to the inverse of the other two. In ToN-IoT, DoS spends fifty six percent of its budget on outgoing bytes, compared with just twelve percent for ransomware in the same dataset.

The pattern is mechanistically legible, which is the part we did not expect. Volumetric attacks are detected by rate, and the cheapest way to disguise one is not to move its volume, which the attack requires, but to stretch the window it is measured over: the policy discovers this without being told, pouring its budget into duration for DoS and Brute Force. Low and slow classes are the opposite case. Bot and Infiltration are already long lived, so duration is not what distinguishes them, and the policy abandons it entirely and pads packet counts instead. It also does not treat identical labels identically, giving ToN-IoT DoS and CSE-CIC DoS opposite allocations, which is correct: those labels come from different capture campaigns and are detected by different features~\cite{sarhan_towards_2022, flood_bad_2024}.

The object an adversary holds after training is therefore not a perturbation. It is a policy that has inferred, from hard labels alone, which feature each class of traffic is detected by, which is a map of the detector's dependence on its own feature space obtained without gradients, without parameters, and without the labels the detector was trained on. It is also more durable than any single perturbation: a defender who blocks a fixed displacement has closed one hole, whereas one who observes this policy has learned only that the adversary knows which hole to use for which attack.

\resultline{The same policy inverts its perturbation strategy across attack classes in one dataset, spending half its budget on flow duration for volumetric attacks and none for low and slow ones. What is learned is a mapping from traffic to weakness, not a constant.}

\subsubsection{Transferability}\label{sec:transfer}

\begin{figure}
    \centering
    \includegraphics[width=\linewidth]{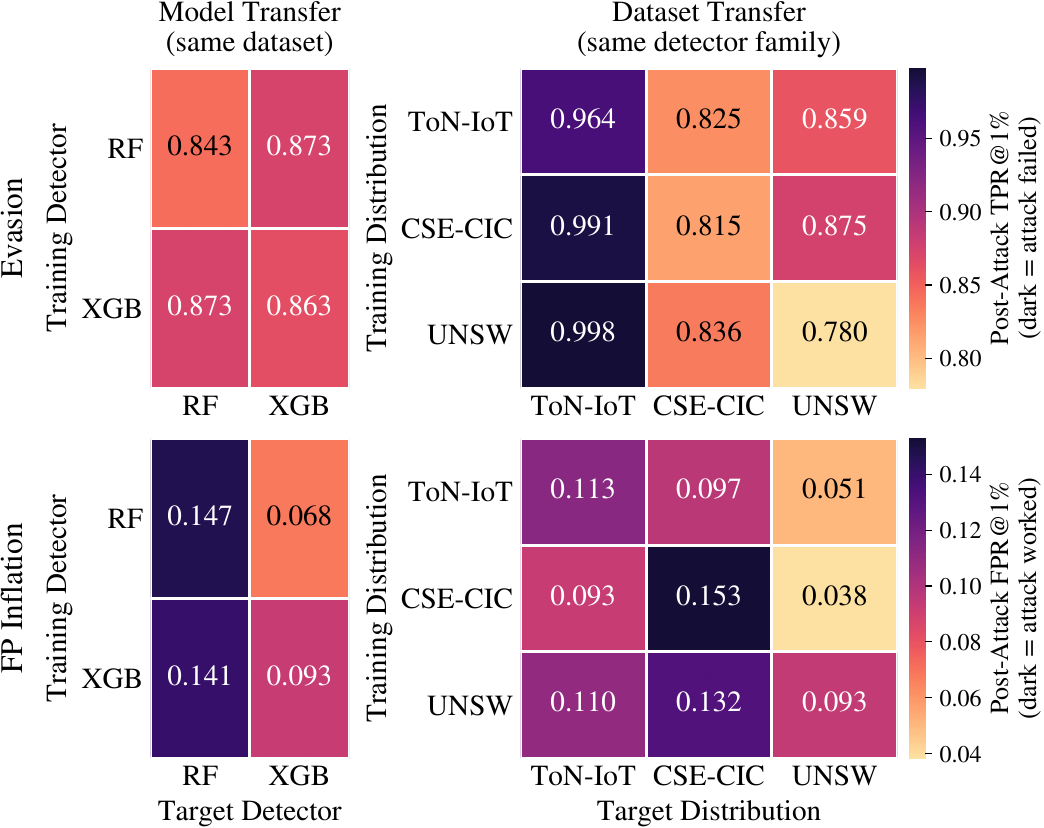}
    \caption{Moving a trained policy to a detector or a distribution it never saw. Rows in each panel are what the policy trained against, columns are what it was deployed against, and the diagonal is the in-distribution case. Top row is retained TPR@1\% (dark = the attack failed), bottom row is the induced FPR (dark = the attack worked).}
    \label{fig:transfer}
\end{figure}

\autoref{sec:threat} argued that the detector an adversary queries is rarely the detector it attacks, and two scenarios make that concrete. In the first, an operator changes classifiers: models are retrained as traffic drifts, architectures are swapped during selection, and a vendor's choice is often opaque to the customer deploying it~\cite{mink_everybodys_2023, apruzzese_sok_2023}, so an adversary that must know the architecture has an attack that expires on the next model release. In the second, and more important, the adversary never had access to the target's traffic at all: it trains on a public NetFlow benchmark and attacks a production network, the cheapest route to a queryable detector and the one an external attacker actually takes. If a policy survives neither move, \autoref{sec:amortization} describes a campaign against exactly one detector on exactly one network, and $N^{*}$ is a far less interesting number than it appears.

We evaluate both. For model transfer we train against one detector family and deploy against the other, holding traffic fixed; for dataset transfer we train on one distribution and deploy against a detector trained on another, holding the family fixed. No additional training queries are spent on the new target. This campaign was run at $\rho = 0.04$, which is not the setting of \autoref{sec:rq1}, so the absolute values should not be read against \autoref{fig:effectiveness}; what matters is the comparison between cells within each panel of \autoref{fig:transfer}.

Model transfer is free. The four evasion cells span $0.843$ to $0.873$ and the diagonal holds no advantage: a policy trained on RandomForest suppresses XGBoost to $0.873$, exactly what an XGBoost trained policy achieves against RandomForest, and the best cell in the panel is not a diagonal at all. Dataset transfer costs more, but not uniformly. Against CSE-CIC, a policy trained on UNSW reaches $0.836$ where one trained on CSE-CIC itself reaches $0.815$, a penalty of two points; against UNSW the penalty is roughly eight.

The surprise is that transfer is the wrong axis. What determines whether the attack works is which detector is attacked, not which one the policy trained on. Every policy fails against ToN-IoT, including the one trained on it, at $0.964$, $0.991$, and $0.998$ down the column against a pre attack rate already near one, and every policy succeeds against CSE-CIC, including the two that never saw it. The columns of \autoref{fig:transfer} are ordered and the rows are barely distinguishable. Mismatch is not what costs the adversary anything; target hardness is, and that is a property of the environment the defender operates in rather than of the adversary's preparation.

The inflation row is stronger still, and it is the result a SOC should read first. All thirteen cells sit above the one percent the detector was calibrated to produce, from $0.038$ at the worst transfer to $0.153$ at the best in distribution, between four and fifteen times the expected alert volume from a policy that in most cells never saw the detector or the traffic. Here too the target detector dominates the outcome: RandomForest is flooded to $0.147$ and $0.141$ regardless of training source, XGBoost only to $0.068$ and $0.093$.

For the economics of \autoref{sec:rq2} this is the decisive point. Search restarts at zero against every new target, so its cost scales with the number of detectors as well as the number of flows. A policy does not: the 81{,}920 queries are spent once and divided across every detector the adversary subsequently attacks, and \autoref{fig:transfer} shows that division costs little for model changes and modestly for distribution changes. An adversary running a campaign across a segmented enterprise, or an entire customer base, reaches break-even proportionally sooner than the single detector figures of \autoref{sec:amortization} report. Amortization holds across targets, not only across inputs.

\resultline{Model transfer is free and dataset transfer costs two to eight points of retained detection. What governs success is which detector is attacked, not which one the policy trained on, and every transfer cell still manufactures four to fifteen times the calibrated alert volume.}

\begin{goal}
\textbf{Answer to \ref{rq:transfer}:} The policy learns a mapping from traffic to detector weakness, not a fixed perturbation, and that mapping survives being moved: changing the detector family costs nothing measurable and changing the traffic distribution costs two to eight points, while every transfer condition still multiplies alert volume at least fourfold. Because one training run covers many targets, transfer lowers the break-even of \autoref{sec:rq2} rather than raising it.
\end{goal}
\section{Related Work \& Discussion}\label{sec:related}

\noindent\textbf{Adversarial examples against ML-NIDS.} Evasion of flow based detectors was established a decade ago~\cite{apruzzese_evading_2018} and refined along one axis since: where the perturbation may be applied. Sheatsley et al. showed that unconstrained feature manipulation produces adversarial examples violating the domain's own semantics, and derived constraints from data to prevent it~\cite{sheatsley_robustness_2021, sheatsley_adversarial_2022}. Pierazzi et al. formalized the feature-space to problem-space gap~\cite{pierazzi_intriguing_2020}, and Verkerken et al. narrowed it, arguing a real adversary perturbs only by issuing commands on hosts it controls~\cite{verkerken_what_2026}. Defenses answer on the same axis: PANTS pairs PGD with an SMT solver to adversarially train on constraint-compliant inputs~\cite{jin_robustifying_2025}. This progression has made the threat model substantially more realistic, but it has left the generation procedure itself largely untouched. This line is therefore orthogonal to ours and complementary: every method above generates perturbations by searching, once per input, and none reports what that search costs at the volume a network produces. We instead hold the perturbation space fixed and measure the generation algorithm itself.

\noindent\textbf{Learned adversaries.} RL has been turned on adversarial generation across security domains: malware detectors~\cite{anderson_evading_2017, tsingenopoulos_how_2024}, CAPTCHA solvers~\cite{tsingenopoulos_captcha_2022}, image classifiers~\cite{domico_adversarial_2026}, and LLM jailbreaking~\cite{chen_when_2024}. Within ML-NIDS the multi-step formulation is the default and is adopted without justification~\cite{apruzzese_deep_2020, wang_crafting_2021, hore_deep_2025}. Closest to us are two network attacks. Amoeba reshapes packet sequences against censorship classifiers and shows a learned policy generalizes to unseen inputs without per-flow optimization~\cite{liu_amoeba_2023}; Deep PackGen learns additive packet perturbations against ML-NIDS and reports strong post-training evasion~\cite{hore_deep_2025}. Both establish that learning works. Neither reports the interactions training consumed, so neither answers whether learning was worth doing: an attack that evades at deployment while spending an unbounded budget beforehand is efficient only under an accounting that omits the budget. This distinction matters because deployment efficiency and end-to-end attack efficiency are not the same quantity. Our contribution is that accounting, and the threshold it yields.

\noindent\textbf{Methodology.} Our framing follows work arguing that ML security evaluations must reflect the conditions under which a system is deployed and attacked. Arp et al. catalogue the pitfalls that inflate reported performance, among them omitting costs a real adversary pays~\cite{arp_dos_2022}. Apruzzese et al. show that real attackers do not compute gradients and interact with systems rather than models~\cite{apruzzese_real_2022}, and that ML-NIDS evaluations routinely report metrics at operating points no analyst deploys~\cite{apruzzese_sok_2023}. McFadden et al. survey RL for cybersecurity and find multi-step formulations adopted without testing whether the problem is sequential, and training costs seldom reported~\cite{mcfadden_sok_2026}. These concerns are coupled in our setting: excluding training cost favors learned attacks, while evaluating at unrealistic operating points can distort both attack effectiveness and the break-even point. We take these as requirements: we calibrate every detector to a stated alert budget, count every query including those spent training, and test the sequential formulation rather than assuming it.

\noindent\textbf{Implications for future work.} Both the adversary's economics and the defender's levers point past the feature space, where our cost model and MDP already operate: only the action set and constraints change under a richer representation. The open extension is the one \autoref{sec:threat} defers, realizing a feature-space target as a host action, a mapping recent work identifies as the unsolved problem~\cite{verkerken_what_2026}. Our contribution sits upstream of it, since knowing which perturbation defeats a detector is a prerequisite for producing one, and we measure the cost of that knowledge. Importantly, moving to the problem space need not change the amortization question; it changes what a policy must output and what constraints it must satisfy. The natural next step reframes generation the way we have reframed it here: pairing a learned policy with a constraint solver over host-realizable packet manipulations, as PANTS does for gradient attacks~\cite{jin_robustifying_2025}, so the amortized object is not a feature-space perturbation but a deployable action.

\section{Conclusion}\label{sec:conclusion}
The role of learning in evading ML-NIDS is amortization. Search retains nothing
between the flows it solves, so its cost grows with the campaign, while a policy
pays once and perturbs cheaply thereafter. We priced both in the same units,
counting every detector query and every second either method spent, training
included, across six ML-NIDS environments at two operating points and under both
adversarial objectives. Break-even arrived after 76 to 1{,}622 adversarial examples, a volume a monitored link produces in seconds of traffic, and past it the policy is cheaper without bound. Along the way we found that the sequential
formulation the literature defaults to is not load bearing, and that what
training buys is a mapping from traffic to detector weakness that holds against
detectors and distributions the policy never queried. The value of learning to attack ML-NIDS was never effectiveness, but the scale at which this domain operates.

\section*{Open Science}\label{sec:open-science}
In accordance with the open science policy, we release the code and artifacts required to reproduce every result in this paper. The repository link at the end of \autoref{sec:intro} contains:

\begin{itemize}
    \item \textbf{Code for the full pipeline of \autoref{fig:overview}}, from detector training and calibration, through policy training and the three search baselines, to the transfer and perturbation-analysis evaluations, with a pinned dependency environment for each stage.
    \item \textbf{Trained policies} for all six environments, both operating points, and both objectives, so the effectiveness, cost, and transfer results can be reproduced without retraining.
    \item \textbf{Per-configuration query and timing logs}, the raw records behind the marginal-cost, break-even, and transfer figures (\autoref{fig:marginal_cost}--\autoref{fig:transfer}), including the single-core timing-probe outputs and their calibration benchmarks (\autoref{timing_env}).
    \item \textbf{Configuration files and plotting scripts} for every experiment, which regenerate the tables and figures of \autoref{sec:evaluation} and the appendix directly from the logs above.
\end{itemize}

The public NetFlow datasets we use~\cite{sarhan_towards_2022} are available from their original sources; our preprocessing, including the deduplication and label-conflict handling of \autoref{data_prep}, is scripted in the repository.

\section*{Ethical Considerations}
This work measures the cost of generating adversarial examples against ML-NIDS, and finds that a learned policy becomes cheaper than search past a break-even volume a network crosses in seconds. The experiments are conducted from an adversary's perspective, but the result's primary use is defensive: it gives operators a concrete lever (\autoref{sec:related}) for raising the cost of attack, and it makes deployment-scale robustness auditing tractable, since an auditor faces the same amortization and enjoys free, unobserved queries against its own detector.

\noindent\textbf{Stakeholders.} The affected parties are operators of deployed ML-NIDS and, by extension, the networks and users those systems protect. Because the paper studies evasion, some findings could in principle help a malicious actor circumvent such a system.

\noindent\textbf{Harms and mitigations.} We identified two potential harms and find both bounded. First, no deployed or commercial system was attacked: every detector in this paper is trained by us on public benchmarks for the experiments, and no third party was involved at any stage. Second, the released artifact (\autoref{sec:open-science}) is a research prototype operating in the feature space; it identifies which perturbation defeats a detector but does not synthesize the host-realizable traffic that would produce one, a gap \autoref{sec:threat} states explicitly and does not close. An actor cannot point it at a production NIDS and obtain a working attack. The transfer result (\autoref{sec:transfer}) shows a trained policy generalizes across detectors, which raises the stakes of that gap but does not bridge it, since the object that transfers is still a feature-space policy.

\noindent\textbf{Decision.} Weighing these, and because the amortization framing and the budget lever it exposes are of greater use to defenders and auditors than the feature-space attack is to an adversary who still lacks a realization step, we proceeded with the research and its publication. We follow responsible-disclosure norms in spirit: the detectors are our own, so there is no third-party system to notify.

\section*{Acknowledgements}
The authors would like to thank Jonathan Gregory for editorial feedback and manuscript improvement. This material is based upon work supported by the National Science Foundation under Grant No. CNS-2343611 and by PRISM, one of seven centers in JUMP 2.0, a Semiconductor Research Corporation (SRC) program sponsored by DARPA. Any opinions, findings, and conclusions or recommendations expressed in this material are those of the author(s) and do not necessarily reflect the views of the National Science Foundation.


\bibliographystyle{plainurl}
\bibliography{refs}

@inproceedings{stateful_carlini,
author = {Chen, Steven and Carlini, Nicholas and Wagner, David},
title = {Stateful Detection of Black-Box Adversarial Attacks},
year = {2020},
isbn = {9781450376112},
publisher = {Association for Computing Machinery},
address = {New York, NY, USA},
url = {https://doi.org/10.1145/3385003.3410925},
doi = {10.1145/3385003.3410925},
booktitle = {Proceedings of the 1st ACM Workshop on Security and Privacy on Artificial Intelligence},
pages = {30–39},
numpages = {10},
location = {Taipei, Taiwan},
series = {SPAI '20}
}

@article{denning_intrusion-detection_1987,
	title = {An {Intrusion}-{Detection} {Model}},
	volume = {SE-13},
	doi = {10.1109/TSE.1987.232894},
	number = {2},
	journal = {IEEE Transactions on Software Engineering},
	author = {Denning, D.E.},
	year = {1987},
	pages = {222--232},
}

@misc{netresec_splitcap_2011,
	title = {{SplitCap}: {A} fast {PCAP} file splitter},
	url = {https://www.netresec.com/?page=SplitCap},
	author = {{Netresec}},
	year = {2011},
}

@misc{edwards_laymans_2025,
	title = {The laymans guide to network performance: {What} are packets per second?},
	url = {https://blog.scsprotect.com/the-laymans-guide-to-network-performance-what-are-packets-per-second},
	author = {Edwards, Devon},
	year = {2025},
	note = {Published: Secure Compliance Solutions Blog},
}

@inproceedings{bhusal_pasa_2024,
	title = {{PASA}: {Attack} {Agnostic} {Unsupervised} {Adversarial} {Detection} {Using} {Prediction} \& {Attribution} {Sensitivity} {Analysis}},
	doi = {10.1109/EuroSP60621.2024.00010},
	booktitle = {2024 {IEEE} 9th {European} {Symposium} on {Security} and {Privacy} ({EuroS}\&{P})},
	author = {Bhusal, Dipkamal and Alam, Md Tanvirul and Veerabhadran, Monish K. and Clifford, Michael and Rampazzi, Sara and Rastogi, Nidhi},
	year = {2024},
	pages = {21--40},
}

@inproceedings{diallo_sabre_2024,
	title = {Sabre: {Cutting} through {Adversarial} {Noise} with {Adaptive} {Spectral} {Filtering} and {Input} {Reconstruction}},
	doi = {10.1109/SP54263.2024.00076},
	booktitle = {2024 {IEEE} {Symposium} on {Security} and {Privacy} ({SP})},
	author = {Diallo, Alec F. and Patras, Paul},
	year = {2024},
	pages = {2901--2919},
}

@misc{cloudflare_one_2020,
	title = {One more ({Zero} {Trust}) thing: {Cloudflare} {Intrusion} {Detection} {System}},
	url = {https://blog.cloudflare.com/one-more-zero-trust-thing-cloudflare-intrusion-detection/},
	author = {{CloudFlare}},
	year = {2020},
}

@techreport{crowley_sans_2025,
	type = {Technical {Report}},
	title = {{SANS} 2025 {SOC} {Survey}},
	url = {https://www.sans.org/white-papers/sans-2025-soc-survey},
	institution = {SANS Research Program},
	author = {Crowley, Christopher},
	year = {2025},
}

@inproceedings{apruzzese_evading_2018,
	title = {Evading {Botnet} {Detectors} based on {Flows} and {Random} {Forest} with {Adversarial} {Samples}},
	booktitle = {2018 {IEEE} 17th {International} {Symposium} on {Network} {Computing} and {Applications} ({NCA})},
	publisher = {IEEE},
	author = {Apruzzese, Giovanni and Colajanni, Michele},
	year = {2018},
	pages = {1--8},
}

@inproceedings{anderson_evading_2017,
	address = {Las Vegas, NV, USA},
	title = {Evading {Machine} {Learning} {Malware} {Detection}},
	url = {https://blackhat.com/docs/us-17/thursday/us-17-Anderson-Bot-Vs-Bot-Evading-Machine-Learning-Malware-Detection-wp.pdf},
	booktitle = {Black {Hat} {USA}},
	author = {Anderson, Hyrum S. and Kharkar, Anant and Filar, Bobby and Roth, Phil},
	month = jul,
	year = {2017},
}

@misc{liu_hard-label_2025,
	title = {A {Hard}-{Label} {Black}-{Box} {Evasion} {Attack} against {ML}-based {Malicious} {Traffic} {Detection} {Systems}},
	url = {https://arxiv.org/abs/2510.14906},
	author = {Liu, Zixuan and Zhao, Yi and Liu, Zhuotao and Li, Qi and Fu, Chuanpu and Zhou, Guangmeng and Xu, Ke},
	year = {2025},
	note = {\_eprint: 2510.14906},
}

@article{chen_when_2024,
	title = {When {LLM} {Meets} {DRL}: {Advancing} {Jailbreaking} {Efficiency} via {DRL}-guided {Search}},
	journal = {arXiv preprint arXiv:2406.08705},
	author = {Chen, Xuan and Nie, Yuzhou and Guo, Wenbo and Zhang, Xiangyu},
	year = {2024},
}

@inproceedings{wang_crafting_2021,
	address = {New York, NY, USA},
	series = {{RAID} '21},
	title = {Crafting {Adversarial} {Example} to {Bypass} {Flow}-\&{ML}- based {Botnet} {Detector} via {RL}},
	isbn = {978-1-4503-9058-3},
	url = {https://doi.org/10.1145/3471621.3471841},
	doi = {10.1145/3471621.3471841},
	booktitle = {Proceedings of the 24th {International} {Symposium} on {Research} in {Attacks}, {Intrusions} and {Defenses}},
	publisher = {Association for Computing Machinery},
	author = {Wang, Junnan and Qixu, Liu and Di, Wu and Dong, Ying and Cui, Xiang},
	year = {2021},
	pages = {193--204},
}

@article{hore_deep_2025,
	address = {New York, NY, USA},
	title = {Deep {PackGen}: {A} {Deep} {Reinforcement} {Learning} {Framework} for {Adversarial} {Network} {Packet} {Generation}},
	volume = {28},
	issn = {2471-2566},
	url = {https://doi.org/10.1145/3712307},
	doi = {10.1145/3712307},
	number = {2},
	journal = {ACM Trans. Priv. Secur.},
	publisher = {Association for Computing Machinery},
	author = {Hore, Soumyadeep and Ghadermazi, Jalal and Paudel, Diwas and Shah, Ankit and Das, Tapas and Bastian, Nathaniel},
	month = feb,
	year = {2025},
}

@article{apruzzese_deep_2020,
	title = {Deep {Reinforcement} {Adversarial} {Learning} {Against} {Botnet} {Evasion} {Attacks}},
	volume = {17},
	doi = {10.1109/TNSM.2020.3031843},
	number = {4},
	journal = {IEEE Transactions on Network and Service Management},
	author = {Apruzzese, Giovanni and Andreolini, Mauro and Marchetti, Mirco and Venturi, Andrea and Colajanni, Michele},
	year = {2020},
	pages = {1975--1987},
}

@inproceedings{domico_adversarial_2026,
	title = {Adversarial {Agents}: {Black}-{Box} {Evasion} {Attacks} with {Reinforcement} {Learning}},
	booktitle = {Proceedings of the {IEEE}/{CVF} {Conference} on {Computer} {Vision} and {Pattern} {Recognition} ({CVPR}) {Findings}},
	author = {Domico, Kyle and Ferrand, Jean-Charles Noirot and Sheatsley, Ryan and Pauley, Eric and Hanna, Josiah and McDaniel, Patrick},
	month = jun,
	year = {2026},
	pages = {647--655},
}

@inproceedings{tsingenopoulos_autoattacker_2019,
	title = {{AutoAttacker}: {A} reinforcement learning approach for black-box adversarial attacks},
	doi = {10.1109/EuroSPW.2019.00032},
	booktitle = {2019 {IEEE} {European} {Symposium} on {Security} and {Privacy} {Workshops} ({EuroS}\&{PW})},
	author = {Tsingenopoulos, Ilias and Preuveneers, Davy and Joosen, Wouter},
	year = {2019},
	pages = {229--237},
}

@inproceedings{tsingenopoulos_captcha_2022,
	title = {Captcha me if you can: {Imitation} {Games} with {Reinforcement} {Learning}},
	doi = {10.1109/EuroSP53844.2022.00050},
	booktitle = {2022 {IEEE} 7th {European} {Symposium} on {Security} and {Privacy} ({EuroS}\&{P})},
	author = {Tsingenopoulos, Ilias and Preuveneers, Davy and Desmet, Lieven and Joosen, Wouter},
	year = {2022},
	pages = {719--735},
}

@inproceedings{tsingenopoulos_how_2024,
	address = {New York, NY, USA},
	series = {{RAID} '24},
	title = {How to {Train} your {Antivirus}: {RL}-based {Hardening} through the {Problem} {Space}},
	isbn = {979-8-4007-0959-3},
	url = {https://doi.org/10.1145/3678890.3678912},
	doi = {10.1145/3678890.3678912},
	booktitle = {Proceedings of the 27th {International} {Symposium} on {Research} in {Attacks}, {Intrusions} and {Defenses}},
	publisher = {Association for Computing Machinery},
	author = {Tsingenopoulos, Ilias and Cortellazzi, Jacopo and Bošanský, Branislav and Aonzo, Simone and Preuveneers, Davy and Joosen, Wouter and Pierazzi, Fabio and Cavallaro, Lorenzo},
	year = {2024},
	pages = {130--146},
}

@inproceedings{tsingenopoulos_adaptive_2025,
	title = {The {Adaptive} {Arms} {Race}: {Redefining} {Robustness} in {AI} {Security}},
	doi = {10.1109/RAID67961.2025.00015},
	booktitle = {2025 28th {International} {Symposium} on {Research} in {Attacks}, {Intrusions} and {Defenses} ({RAID})},
	author = {Tsingenopoulos, Ilias and Rimmer, Vera and Preuveneers, Davy and Pierazzi, Fabio and Cavallaro, Lorenzo and Joosen, Wouter},
	year = {2025},
	pages = {611--628},
}

@misc{mcfadden_sok_2026,
	title = {{SoK}: {The} {Pitfalls} of {Deep} {Reinforcement} {Learning} for {Cybersecurity}},
	url = {https://arxiv.org/abs/2602.08690},
	author = {McFadden, Shae and Foley, Myles and Bates, Elizabeth and Tsingenopoulos, Ilias and Vyas, Sanyam and Mavroudis, Vasilios and Hicks, Chris and Pierazzi, Fabio},
	year = {2026},
	note = {\_eprint: 2602.08690},
}

@inproceedings{wang_manda_2021,
	title = {{MANDA}: {On} {Adversarial} {Example} {Detection} for {Network} {Intrusion} {Detection} {System}},
	doi = {10.1109/INFOCOM42981.2021.9488874},
	booktitle = {{IEEE} {INFOCOM} 2021 - {IEEE} {Conference} on {Computer} {Communications}},
	author = {Wang, Ning and Chen, Yimin and Hu, Yang and Lou, Wenjing and Hou, Y. Thomas},
	year = {2021},
	pages = {1--10},
}

@inproceedings{pierazzi_intriguing_2020,
	title = {Intriguing {Properties} of {Adversarial} {ML} {Attacks} in the {Problem} {Space}},
	doi = {10.1109/SP40000.2020.00073},
	booktitle = {2020 {IEEE} {Symposium} on {Security} and {Privacy} ({SP})},
	author = {Pierazzi, Fabio and Pendlebury, Feargus and Cortellazzi, Jacopo and Cavallaro, Lorenzo},
	year = {2020},
	pages = {1332--1349},
}

@inproceedings{verkerken_what_2026,
	address = {New York, NY, USA},
	series = {{ASIA} {CCS} '26},
	title = {“{What} is the {Problem} {Space}?” {Defining} {Host}-space {Adversarial} {Perturbations} against {Network} {Intrusion} {Detection} {Systems}},
	isbn = {979-8-4007-2356-8},
	url = {https://doi.org/10.1145/3779208.3807482},
	doi = {10.1145/3779208.3807482},
	booktitle = {Proceedings of the {ACM} {Asia} {Conference} on {Computer} and {Communications} {Security}},
	publisher = {Association for Computing Machinery},
	author = {Verkerken, Miel and D'hooge, Laurens and Volckaert, Bruno and De Turck, Filip and Apruzzese, Giovanni},
	year = {2026},
	pages = {1043--1059},
}

@article{liu_amoeba_2023,
	address = {New York, NY, USA},
	title = {Amoeba: {Circumventing} {ML}-supported {Network} {Censorship} via {Adversarial} {Reinforcement} {Learning}},
	volume = {1},
	url = {https://doi.org/10.1145/3629131},
	doi = {10.1145/3629131},
	number = {CoNEXT3},
	journal = {Proc. ACM Netw.},
	publisher = {Association for Computing Machinery},
	author = {Liu, Haoyu and Diallo, Alec F. and Patras, Paul},
	month = nov,
	year = {2023},
}

@inproceedings{sheatsley_robustness_2021,
	address = {New York, NY, USA},
	series = {{CCS} '21},
	title = {On the {Robustness} of {Domain} {Constraints}},
	isbn = {978-1-4503-8454-4},
	url = {https://doi.org/10.1145/3460120.3484570},
	doi = {10.1145/3460120.3484570},
	booktitle = {Proceedings of the 2021 {ACM} {SIGSAC} {Conference} on {Computer} and {Communications} {Security}},
	publisher = {Association for Computing Machinery},
	author = {Sheatsley, Ryan and Hoak, Blaine and Pauley, Eric and Beugin, Yohan and Weisman, Michael J. and McDaniel, Patrick},
	year = {2021},
	pages = {495--515},
}

@techreport{ptacek_insertion_1998,
    title = {Insertion, evasion, and denial of service: {Eluding} network intrusion detection},
    author = {Ptacek, Thomas H and Newsham, Timothy N},
    year = {1998},
    institution = {Secure Networks, Inc.},
    month = jan
}

@inproceedings{jin_robustifying_2025,
	address = {Seattle, WA},
	title = {Robustifying {ML}-powered {Network} {Classifiers} with {PANTS}},
	isbn = {978-1-939133-52-6},
	url = {https://www.usenix.org/conference/usenixsecurity25/presentation/jin-minhao},
	booktitle = {34th {USENIX} {Security} {Symposium} ({USENIX} {Security} 25)},
	publisher = {USENIX Association},
	author = {Jin, Minhao and Apostolaki, Maria},
	month = aug,
	year = {2025},
	pages = {7291--7310},
}

@article{tariq_alert_2025,
	address = {New York, NY, USA},
	title = {Alert {Fatigue} in {Security} {Operations} {Centres}: {Research} {Challenges} and {Opportunities}},
	volume = {57},
	issn = {0360-0300},
	url = {https://doi.org/10.1145/3723158},
	doi = {10.1145/3723158},
	number = {9},
	journal = {ACM Comput. Surv.},
	publisher = {Association for Computing Machinery},
	author = {Tariq, Shahroz and Baruwal Chhetri, Mohan and Nepal, Surya and Paris, Cecile},
	month = apr,
	year = {2025},
}

@misc{guo_knowml_2026,
	title = {{KnowML}: {Improving} {Generalization} of {ML}-{NIDS} with {Attack} {Knowledge} {Graphs}},
	url = {https://arxiv.org/abs/2506.19802},
	author = {Guo, Xin Fan and Zheng, Xinran and Penuela, Albert Merono and Cavallaro, Lorenzo and Maffeis, Sergio and Pierazzi, Fabio},
	year = {2026},
	note = {\_eprint: 2506.19802},
}

@inproceedings{ede_deepcase_2022,
	title = {{DEEPCASE}: {Semi}-{Supervised} {Contextual} {Analysis} of {Security} {Events}},
	doi = {10.1109/SP46214.2022.9833671},
	booktitle = {2022 {IEEE} {Symposium} on {Security} and {Privacy} ({SP})},
	author = {Ede, Thijs van and Aghakhani, Hojjat and Spahn, Noah and Bortolameotti, Riccardo and Cova, Marco and Continella, Andrea and Steen, Maarten van and Peter, Andreas and Kruegel, Christopher and Vigna, Giovanni},
	year = {2022},
	pages = {522--539},
}

@inproceedings{engelen_troubleshooting_2021,
	title = {Troubleshooting an {Intrusion} {Detection} {Dataset}: the {CICIDS2017} {Case} {Study}},
	doi = {10.1109/SPW53761.2021.00009},
	booktitle = {2021 {IEEE} {Security} and {Privacy} {Workshops} ({SPW})},
	author = {Engelen, Gints and Rimmer, Vera and Joosen, Wouter},
	year = {2021},
	pages = {7--12},
}

@inproceedings{mink_everybodys_2023,
	title = {Everybody’s {Got} {ML}, {Tell} {Me} {What} {Else} {You} {Have}: {Practitioners}’ {Perception} of {ML}-{Based} {Security} {Tools} and {Explanations}},
	doi = {10.1109/SP46215.2023.10179321},
	booktitle = {2023 {IEEE} {Symposium} on {Security} and {Privacy} ({SP})},
	author = {Mink, Jaron and Benkraouda, Hadjer and Yang, Limin and Ciptadi, Arridhana and Ahmadzadeh, Ali and Votipka, Daniel and Wang, Gang},
	year = {2023},
	pages = {2068--2085},
}

@inproceedings{alahmadi_99_2022,
	address = {Boston, MA},
	title = {99\% {False} {Positives}: {A} {Qualitative} {Study} of {SOC} {Analysts}’ {Perspectives} on {Security} {Alarms}},
	isbn = {978-1-939133-31-1},
	url = {https://www.usenix.org/conference/usenixsecurity22/presentation/alahmadi},
	booktitle = {31st {USENIX} {Security} {Symposium} ({USENIX} {Security} 22)},
	publisher = {USENIX Association},
	author = {Alahmadi, Bushra A. and Axon, Louise and Martinovic, Ivan},
	month = aug,
	year = {2022},
	pages = {2783--2800},
}

@inproceedings{barradas_flowlens_2021,
	address = {San Diego, CA, USA},
	title = {{FlowLens}: {Enabling} {Efficient} {Flow} {Classification} for {ML}-based {Network} {Security} {Applications}},
	booktitle = {Proceedings of the 28th {Network} and {Distributed} {System} {Security} {Symposium}},
	author = {Barradas, Diogo and Santos, Nuno and Rodrigues, Luís and Signorello, Salvatore and Ramos, Fernando M. V. and Madeira, André},
	year = {2021},
}

@inproceedings{apruzzese_sok_2023,
	title = {{SoK}: {Pragmatic} {Assessment} of {Machine} {Learning} for {Network} {Intrusion} {Detection}},
	booktitle = {8th {IEEE} {European} {Symposium} on {Security} and {Privacy} ({EuroSP})},
	publisher = {IEEE},
	author = {Apruzzese, Giovanni and Laskov, Pavel and Schneider, Johannes},
	year = {2023},
}

@article{han_evaluating_2021,
	title = {Evaluating and {Improving} {Adversarial} {Robustness} of {Machine} {Learning}-{Based} {Network} {Intrusion} {Detectors}},
	volume = {39},
	doi = {10.1109/JSAC.2021.3087242},
	number = {8},
	journal = {IEEE Journal on Selected Areas in Communications},
	author = {Han, Dongqi and Wang, Zhiliang and Zhong, Ying and Chen, Wenqi and Yang, Jiahai and Lu, Shuqiang and Shi, Xingang and Yin, Xia},
	year = {2021},
	pages = {2632--2647},
}

@inproceedings{flood_bad_2024,
	title = {Bad {Design} {Smells} in {Benchmark} {NIDS} {Datasets}},
	doi = {10.1109/EuroSP60621.2024.00042},
	booktitle = {2024 {IEEE} 9th {European} {Symposium} on {Security} and {Privacy} ({EuroS}\&{P})},
	author = {Flood, Robert and Engelen, Gints and Aspinall, David and Desmet, Lieven},
	year = {2024},
	pages = {658--675},
}

@inproceedings{apruzzese_sok_2026,
	address = {New York, NY, USA},
	series = {{ASIA} {CCS} '26},
	title = {{SoK}: {Reshaping} {Research} on {Network} {Intrusion} {Detection} {Systems}},
	isbn = {979-8-4007-2356-8},
	url = {https://doi.org/10.1145/3779208.3807488},
	doi = {10.1145/3779208.3807488},
	booktitle = {Proceedings of the {ACM} {Asia} {Conference} on {Computer} and {Communications} {Security}},
	publisher = {Association for Computing Machinery},
	author = {Apruzzese, Giovanni},
	year = {2026},
	pages = {1028--1042},
}

@article{layeghy_benchmarking_2024,
	title = {Benchmarking the benchmark — {Comparing} synthetic and real-world {Network} {IDS} datasets},
	volume = {80},
	issn = {2214-2126},
	url = {https://www.sciencedirect.com/science/article/pii/S2214212623002739},
	doi = {https://doi.org/10.1016/j.jisa.2023.103689},
	journal = {Journal of Information Security and Applications},
	author = {Layeghy, Siamak and Gallagher, Marcus and Portmann, Marius},
	year = {2024},
	pages = {103689},
}

@techreport{claise_cisco_2004,
	type = {{RFC}},
	title = {Cisco {Systems} {NetFlow} {Services} {Export} {Version} 9},
	url = {https://datatracker.ietf.org/doc/html/rfc3954},
	number = {3954},
	institution = {Internet Engineering Task Force},
	author = {Claise, Benoît},
	month = oct,
	year = {2004},
	note = {Backup Publisher: Internet Engineering Task Force
Published: Internet RFC 3954},
}

@inproceedings{anderson_practical_2021,
    title = {The {Practical} {Divide} between {Adversarial} {ML} {Research} and {Security} {Practice}: {A} {Red} {Team} {Perspective}},
    booktitle = {2021 {USENIX} {Enigma} {Conference} ({Enigma} 21)},
    publisher = {USENIX Association},
    author = {Anderson, Hyrum},
    month = feb,
    year = {2021},
}

@article{merkle_economics_2024,
	title = {On the {Economics} of {Adversarial} {Machine} {Learning}},
	volume = {19},
	issn = {1556-6013},
	url = {https://doi.org/10.1109/TIFS.2024.3379829},
	doi = {10.1109/TIFS.2024.3379829},
	journal = {Trans. Info. For. Sec.},
	publisher = {IEEE Press},
	author = {Merkle, Florian and Samsinger, Maximilian and Schöttle, Pascal and Pevny, Tomas},
	month = jan,
	year = {2024},
	pages = {4670--4685},
}

@inproceedings {nasr_defeating_2021,
author = {Milad Nasr and Alireza Bahramali and Amir Houmansadr},
title = {Defeating {DNN-Based} Traffic Analysis Systems in {Real-Time} With Blind Adversarial Perturbations},
booktitle = {30th USENIX Security Symposium (USENIX Security 21)},
year = {2021},
isbn = {978-1-939133-24-3},
pages = {2705--2722},
url = {https://www.usenix.org/conference/usenixsecurity21/presentation/nasr},
publisher = {USENIX Association},
month = aug
}

@inproceedings {arp_dos_2022,
author = {Daniel Arp and Erwin Quiring and Feargus Pendlebury and Alexander Warnecke and Fabio Pierazzi and Christian Wressnegger and Lorenzo Cavallaro and Konrad Rieck},
title = {Dos and Don{\textquoteright}ts of Machine Learning in Computer Security},
booktitle = {31st USENIX Security Symposium (USENIX Security 22)},
year = {2022},
isbn = {978-1-939133-31-1},
address = {Boston, MA},
pages = {3971--3988},
url = {https://www.usenix.org/conference/usenixsecurity22/presentation/arp},
publisher = {USENIX Association},
month = aug
}

@inproceedings{sarhan_netflow_2021,
	address = {Cham},
	title = {{NetFlow} {Datasets} for {Machine} {Learning}-{Based} {Network} {Intrusion} {Detection} {Systems}},
	isbn = {978-3-030-72802-1},
	booktitle = {Big {Data} {Technologies} and {Applications}},
	publisher = {Springer International Publishing},
	author = {Sarhan, Mohanad and Layeghy, Siamak and Moustafa, Nour and Portmann, Marius},
	editor = {Deze, Zeng and Huang, Huan and Hou, Rui and Rho, Seungmin and Chilamkurti, Naveen},
	year = {2021},
	pages = {117--135},
}

@misc{mnih_asynchronous_2016,
	title = {Asynchronous {Methods} for {Deep} {Reinforcement} {Learning}},
	url = {http://arxiv.org/abs/1602.01783},
	doi = {10.48550/arXiv.1602.01783},
	urldate = {2026-02-02},
	publisher = {arXiv},
	author = {Mnih, Volodymyr and Badia, Adrià Puigdomènech and Mirza, Mehdi and Graves, Alex and Lillicrap, Timothy P. and Harley, Tim and Silver, David and Kavukcuoglu, Koray},
	month = jun,
	year = {2016},
	note = {arXiv:1602.01783 [cs]},
}

@article{sharafaldin_toward_2018,
	title = {Toward generating a new intrusion detection dataset and intrusion traffic characterization.},
	volume = {1},
	number = {2018},
	journal = {ICISSp},
	author = {Sharafaldin, Iman and Lashkari, Arash Habibi and Ghorbani, Ali A and {others}},
	year = {2018},
	pages = {108--116},
}

@article{alsaedi_ton_iot_2020,
	title = {{TON}\_IoT {Telemetry} {Dataset}: {A} {New} {Generation} {Dataset} of {IoT} and {IIoT} for {Data}-{Driven} {Intrusion} {Detection} {Systems}},
	volume = {8},
	issn = {2169-3536},
	shorttitle = {{TON}\_IoT {Telemetry} {Dataset}},
	url = {https://ieeexplore.ieee.org/document/9189760},
	doi = {10.1109/ACCESS.2020.3022862},
	urldate = {2025-11-19},
	journal = {IEEE Access},
	author = {Alsaedi, Abdullah and Moustafa, Nour and Tari, Zahir and Mahmood, Abdun and Anwar, Adnan},
	year = {2020},
	pages = {165130--165150},
}

@inproceedings{moustafa_unsw-nb15_2015,
	title = {{UNSW}-{NB15}: a comprehensive data set for network intrusion detection systems ({UNSW}-{NB15} network data set)},
	shorttitle = {{UNSW}-{NB15}},
	url = {https://ieeexplore.ieee.org/document/7348942},
	doi = {10.1109/MilCIS.2015.7348942},
	urldate = {2025-11-19},
	booktitle = {2015 {Military} {Communications} and {Information} {Systems} {Conference} ({MilCIS})},
	author = {Moustafa, Nour and Slay, Jill},
	month = nov,
	year = {2015},
	pages = {1--6},
}

@article{sarhan_towards_2022,
	title = {Towards a {Standard} {Feature} {Set} for {Network} {Intrusion} {Detection} {System} {Datasets}},
	volume = {27},
	issn = {1572-8153},
	url = {https://doi.org/10.1007/s11036-021-01843-0},
	doi = {10.1007/s11036-021-01843-0},
	language = {en},
	number = {1},
	urldate = {2025-11-19},
	journal = {Mobile Networks and Applications},
	author = {Sarhan, Mohanad and Layeghy, Siamak and Portmann, Marius},
	month = feb,
	year = {2022},
	pages = {357--370},
}

@misc{goodfellow_explaining_2015,
	title = {Explaining and {Harnessing} {Adversarial} {Examples}},
	url = {http://arxiv.org/abs/1412.6572},
	doi = {10.48550/arXiv.1412.6572},
	urldate = {2025-10-22},
	publisher = {arXiv},
	author = {Goodfellow, Ian J. and Shlens, Jonathon and Szegedy, Christian},
	month = mar,
	year = {2015},
	note = {arXiv:1412.6572 [stat]},
}

@misc{apruzzese_real_2022,
	title = {"{Real} {Attackers} {Don}'t {Compute} {Gradients}": {Bridging} the {Gap} {Between} {Adversarial} {ML} {Research} and {Practice}},
	shorttitle = {"{Real} {Attackers} {Don}'t {Compute} {Gradients}"},
	url = {http://arxiv.org/abs/2212.14315},
	doi = {10.48550/arXiv.2212.14315},
	urldate = {2025-09-23},
	publisher = {arXiv},
	author = {Apruzzese, Giovanni and Anderson, Hyrum S. and Dambra, Savino and Freeman, David and Pierazzi, Fabio and Roundy, Kevin A.},
	month = dec,
	year = {2022},
	note = {arXiv:2212.14315 [cs]},
}

@misc{sheatsley_adversarial_2022,
	title = {Adversarial {Examples} in {Constrained} {Domains}},
	url = {http://arxiv.org/abs/2011.01183},
	doi = {10.48550/arXiv.2011.01183},
	urldate = {2025-09-11},
	publisher = {arXiv},
	author = {Sheatsley, Ryan and Papernot, Nicolas and Weisman, Michael and Verma, Gunjan and McDaniel, Patrick},
	month = sep,
	year = {2022},
	note = {arXiv:2011.01183 [cs]},
}

@misc{szegedy_intriguing_2014,
	title = {Intriguing properties of neural networks},
	url = {http://arxiv.org/abs/1312.6199},
	doi = {10.48550/arXiv.1312.6199},
	urldate = {2025-08-26},
	publisher = {arXiv},
	author = {Szegedy, Christian and Zaremba, Wojciech and Sutskever, Ilya and Bruna, Joan and Erhan, Dumitru and Goodfellow, Ian and Fergus, Rob},
	month = feb,
	year = {2014},
	note = {arXiv:1312.6199 [cs]},
}

@misc{chen_hopskipjumpattack_2020,
	title = {{HopSkipJumpAttack}: {A} {Query}-{Efficient} {Decision}-{Based} {Attack}},
	shorttitle = {{HopSkipJumpAttack}},
	url = {http://arxiv.org/abs/1904.02144},
	doi = {10.48550/arXiv.1904.02144},
	urldate = {2025-05-30},
	publisher = {arXiv},
	author = {Chen, Jianbo and Jordan, Michael I. and Wainwright, Martin J.},
	month = apr,
	year = {2020},
	note = {arXiv:1904.02144 [cs]},
}

@misc{madry_towards_2019,
	title = {Towards {Deep} {Learning} {Models} {Resistant} to {Adversarial} {Attacks}},
	url = {http://arxiv.org/abs/1706.06083},
	doi = {10.48550/arXiv.1706.06083},
	urldate = {2025-05-30},
	publisher = {arXiv},
	author = {Madry, Aleksander and Makelov, Aleksandar and Schmidt, Ludwig and Tsipras, Dimitris and Vladu, Adrian},
	month = sep,
	year = {2019},
	note = {arXiv:1706.06083 [stat]},
}

@book{sutton_reinforcement_2020,
	address = {Cambridge, Massachusetts London, England},
	edition = {Second edition},
	series = {Adaptive computation and machine learning},
	title = {Reinforcement learning: an introduction},
	isbn = {978-0-262-03924-6},
	shorttitle = {Reinforcement learning},
	language = {en},
	publisher = {The MIT Press},
	author = {Sutton, Richard S. and Barto, Andrew},
	year = {2020},
}

@misc{haarnoja_soft_2018,
	title = {Soft {Actor}-{Critic}: {Off}-{Policy} {Maximum} {Entropy} {Deep} {Reinforcement} {Learning} with a {Stochastic} {Actor}},
	shorttitle = {Soft {Actor}-{Critic}},
	url = {http://arxiv.org/abs/1801.01290},
	doi = {10.48550/arXiv.1801.01290},
	urldate = {2025-03-19},
	publisher = {arXiv},
	author = {Haarnoja, Tuomas and Zhou, Aurick and Abbeel, Pieter and Levine, Sergey},
	month = aug,
	year = {2018},
	note = {arXiv:1801.01290 [cs]},
}

@misc{schulman_proximal_2017,
	title = {Proximal {Policy} {Optimization} {Algorithms}},
	url = {http://arxiv.org/abs/1707.06347},
	doi = {10.48550/arXiv.1707.06347},
	urldate = {2025-03-19},
	publisher = {arXiv},
	author = {Schulman, John and Wolski, Filip and Dhariwal, Prafulla and Radford, Alec and Klimov, Oleg},
	month = aug,
	year = {2017},
	note = {arXiv:1707.06347 [cs]},
}

@inproceedings{biggio_evasion_2013,
	address = {Berlin, Heidelberg},
	title = {Evasion {Attacks} against {Machine} {Learning} at {Test} {Time}},
	isbn = {978-3-642-40994-3},
	doi = {10.1007/978-3-642-40994-3_25},
	language = {en},
	booktitle = {Machine {Learning} and {Knowledge} {Discovery} in {Databases}},
	publisher = {Springer},
	author = {Biggio, Battista and Corona, Igino and Maiorca, Davide and Nelson, Blaine and Šrndić, Nedim and Laskov, Pavel and Giacinto, Giorgio and Roli, Fabio},
	editor = {Blockeel, Hendrik and Kersting, Kristian and Nijssen, Siegfried and Železný, Filip},
	year = {2013},
	pages = {387--402},
}

@inproceedings{moosavi-dezfooli_universal_2017,
	address = {Honolulu, HI},
	title = {Universal {Adversarial} {Perturbations}},
	isbn = {978-1-5386-0457-1},
	url = {http://ieeexplore.ieee.org/document/8099500/},
	doi = {10.1109/CVPR.2017.17},
	language = {en},
	urldate = {2025-02-10},
	booktitle = {2017 {IEEE} {Conference} on {Computer} {Vision} and {Pattern} {Recognition} ({CVPR})},
	publisher = {IEEE},
	author = {Moosavi-Dezfooli, Seyed-Mohsen and Fawzi, Alhussein and Fawzi, Omar and Frossard, Pascal},
	month = jul,
	year = {2017},
	pages = {86--94},
}

@article{madry_towards_2017,
	title = {Towards deep learning models resistant to adversarial attacks},
	journal = {arXiv preprint arXiv:1706.06083},
	author = {Madry, Aleksander},
	year = {2017},
}

@article{raffin_stable-baselines3_2021,
	title = {Stable-{Baselines3}: {Reliable} {Reinforcement} {Learning} {Implementations}},
	volume = {22},
	issn = {1533-7928},
	shorttitle = {Stable-{Baselines3}},
	url = {http://jmlr.org/papers/v22/20-1364.html},
	number = {268},
	urldate = {2024-07-29},
	journal = {Journal of Machine Learning Research},
	author = {Raffin, Antonin and Hill, Ashley and Gleave, Adam and Kanervisto, Anssi and Ernestus, Maximilian and Dormann, Noah},
	year = {2021},
	pages = {1--8},
}

@inproceedings{papernot_limitations_2016,
	title = {The {Limitations} of {Deep} {Learning} in {Adversarial} {Settings}},
	url = {https://ieeexplore.ieee.org/abstract/document/7467366},
	doi = {10.1109/EuroSP.2016.36},
	urldate = {2024-07-29},
	booktitle = {2016 {IEEE} {European} {Symposium} on {Security} and {Privacy} ({EuroS}\&{P})},
	author = {Papernot, Nicolas and McDaniel, Patrick and Jha, Somesh and Fredrikson, Matt and Celik, Z. Berkay and Swami, Ananthram},
	month = mar,
	year = {2016},
	pages = {372--387},
}

\section*{Generative AI Usage Considerations}

Generative AI was used to assist with Python implementation, primarily for code refactoring, debugging, and plotting scripts. Generative AI was also used for editorial purposes in this manuscript, and all outputs were inspected by the authors to ensure accuracy and originality. The research questions, metrics, experimental design, analysis, and technical claims were developed and validated by the authors.

\appendix

\section{Dataset Preparation and Attack Distribution}\label{data_prep}
\autoref{tab:attack_dist_table} reports the per-class flow counts for each of the three datasets after preprocessing, spanning the 21 attack classes evaluated in \autoref{sec:universal}. All three datasets are known to carry label and duplicate defects~\cite{engelen_troubleshooting_2021, flood_bad_2024}. We deduplicate on the ten model features of \autoref{tab:features} and, where deduplication leaves records with conflicting labels, retain them rather than resolve them arbitrarily. This admits a residual population of feature-identical flows with disagreeing labels, which upper-bounds the true positive rate any detector can achieve on that dataset regardless of attack. \autoref{tab:tpr_ceiling} reports this ceiling per dataset; every pre-attack TPR in \autoref{tab:detectors} sits at or below it.

\begin{table}[t]
\centering
\setlength{\tabcolsep}{4pt}
\footnotesize
\begin{tabular}{lrrr}
\toprule
Class & ToN-IoT & UNSW-NB15 & CSE-CIC-IDS2018 \\
\midrule
Benign & 198{,}450 & 1{,}421{,}093 & 5{,}534{,}773 \\
\midrule
\multicolumn{4}{l}{\itshape ToN-IoT} \\
DoS & 15{,}021 & 3{,}277 & 171{,}936 \\
scanning & 19{,}110 & -- & -- \\
mitm & 794 & -- & -- \\
DDoS & 106{,}145 & -- & 51{,}418 \\
injection & 445{,}636 & -- & 31 \\
Backdoor & 17{,}219 & 317 & -- \\
ransomware & 142 & -- & -- \\
\addlinespace
\multicolumn{4}{l}{\itshape UNSW-NB15} \\
Exploits & -- & 22{,}830 & -- \\
Fuzzers & -- & 17{,}994 & -- \\
Generic & -- & 3{,}628 & -- \\
Shellcode & -- & 1{,}365 & -- \\
Worms & -- & 142 & -- \\
Reconnaissance & -- & 10{,}395 & -- \\
Analysis & -- & 382 & -- \\
\addlinespace
\multicolumn{4}{l}{\itshape CSE-CIC-IDS2018} \\
Infiltration & -- & -- & 59{,}373 \\
Brute Force & -- & -- & 71{,}363 \\
Bot & -- & -- & 15{,}498 \\
\midrule
Total & 802{,}517 & 1{,}481{,}423 & 5{,}904{,}392 \\
\bottomrule
\end{tabular}

\caption{Flow counts per attack class across the three datasets after deduplication. Benign counts are listed separately from the 21 attack classes.}
\label{tab:attack_dist_table}
\end{table}

\begin{table}[t]
\centering
\setlength{\tabcolsep}{4pt}
\small
\begin{tabular}{lrrrr}
\toprule
 & \multicolumn{2}{c}{Conflicting-label} & \multicolumn{2}{c}{TPR ceiling} \\
\cmidrule(lr){2-3}\cmidrule(lr){4-5}
Dataset & groups & flows & $\phi{=}1\%$ & $\phi{=}0.1\%$ \\
\midrule
ToN-IoT & 2 & 4 & 1.0000 & 1.0000 \\
CSE-CIC-IDS2018 & 7{,}649 & 15{,}298 & 1.0000 & 0.9943 \\
UNSW-NB15 & 149 & 298 & 1.0000 & 1.0000 \\
\bottomrule
\end{tabular}

\caption{Per-dataset TPR ceiling induced by feature-identical flows with conflicting labels after deduplication. No detector calibrated on the dataset can exceed this rate, and \autoref{tab:detectors} is bounded by it.}
\label{tab:tpr_ceiling}
\end{table}

\section{Detector Selection and Calibration}\label{detector_sweep}
Detector hyperparameters are selected by TPR at the target FPR on a validation split, over the grids in \autoref{tab:detector_hparams}. \autoref{tab:calibration} then reports, for each of the six environments and both operating points, the calibrated threshold $\tau_\varphi$, the achieved FPR on the benign test holdout, and the benign holdout size. The two cells marked in \autoref{tab:detectors} are ToN-IoT at $\varphi = 0.1\%$, where the benign holdout supports the threshold only marginally; \autoref{tab:calibration} gives the holdout size that motivates that caution.

\begin{table}[t]
\centering
\setlength{\tabcolsep}{4pt}
\footnotesize
\begin{tabular}{@{}llrrr@{}}
\toprule
Hyperparameter & Grid & ToN-IoT & UNSW & CSE-CIC \\
\midrule
\multicolumn{5}{@{}l}{\itshape Random forest} \\
n\_estimators    & $\{200, 500\}$          & 200  & 200  & 200 \\
max\_depth       & $\{\textsc{none}, 30\}$ & \textsc{none} & \textsc{none} & 30 \\
min\_samples\_leaf & $\{1, 10\}$           & 1    & 10   & 1 \\
\addlinespace
\multicolumn{5}{@{}l}{\itshape XGBoost} \\
n\_estimators  & $\{300, 800\}$      & 300  & 300  & 300 \\
max\_depth     & $\{6, 10\}$         & 10   & 10   & 10 \\
learning\_rate & $\{0.05, 0.15\}$    & 0.05 & 0.05 & 0.05 \\
subsample      & ---                 & 1.0  & 1.0  & 1.0 \\
\bottomrule
\end{tabular}

\caption{Hyperparameter grids for RF and XGB, and the configuration selected per dataset by TPR at the target FPR.}
\label{tab:detector_hparams}
\end{table}

\begin{table}[t]
\centering
\setlength{\tabcolsep}{3pt}
\footnotesize
\begin{tabular}{llrrrrr}
\toprule
 & & \multicolumn{2}{c}{$\phi = 1\%$} & \multicolumn{2}{c}{$\phi = 0.1\%$} & Benign \\
\cmidrule(lr){3-4}\cmidrule(lr){5-6}
Dataset & Det. & $\tau$ & test FPR & $\tau$ & test FPR & holdout \\
\midrule
ToN-IoT & RF & 0.1000 & 1.013\% & 0.7481 & 0.103\%$^{\dagger}$ & 39{,}690 \\
 & XGB & 0.0476 & 0.978\% & 0.8729 & 0.123\%$^{\dagger}$ & 39{,}690 \\
\addlinespace
UNSW & RF & 0.3768 & 0.970\% & 0.8327 & 0.100\% & 284{,}219 \\
 & XGB & 0.3331 & 0.980\% & 0.9267 & 0.098\% & 284{,}219 \\
\addlinespace
CSE-CIC & RF & 0.1106 & 1.004\% & 0.3592 & 0.101\% & 1{,}106{,}955 \\
 & XGB & 0.0446 & 1.001\% & 0.1249 & 0.101\% & 1{,}106{,}955 \\
\bottomrule
\end{tabular}

\caption{Calibrated threshold $\tau_\varphi$, achieved test FPR, and benign holdout size for each environment at $\varphi = 1\%$ and $\varphi = 0.1\%$. $^{\dagger}$ marks the marginally-supported ToN-IoT cells of \autoref{tab:detectors}.}
\label{tab:calibration}
\end{table}

\section{Baseline and Policy Configurations}\label{baseline_optimization}
Every result compares methods under a shared tuning budget of 1000 queries per method (\autoref{sec:cost-model}), and both the search baselines and the policy are tuned within it. \autoref{tab:baseline_config} lists the tuned configuration of PGD, HSJA, and random search: step size, iteration cap, and, for PGD, the surrogate architecture and the fraction of queries that reach the surrogate rather than the victim. \autoref{tab:policy_config} lists the policy configuration: the actor-critic architecture, the training length of $8\times10^4$ environment steps, and the per-algorithm hyperparameters for PPO, SAC, and A2C. Because $N^*$ is a ratio whose denominator is a difference of measured marginal costs (\autoref{sec:cost-model}), it is sensitive to baseline tuning; the configurations here are the ones behind every $N^*$ we report.

\begin{table}[t]
\centering
\setlength{\tabcolsep}{4pt}
\footnotesize
\begin{tabular}{@{}llp{3.3cm}l@{}}
\toprule
Method & Step size & Tuned settings & Surrogate \\
\midrule
PGD & $0.02\rho$ & 200 steps, 5 restarts, confirm every 5 & MLP $[256, 128]$ \\
\addlinespace
HSJA & $\propto \lVert\delta\rVert/\sqrt{t}$ & 25 iters, 100 init + 10 gradient probes & --- \\
\addlinespace
Random & --- & uniform in box, batch 256 & --- \\
\midrule
\multicolumn{4}{@{}p{\dimexpr\columnwidth-2\tabcolsep}@{}}{Query cap 1{,}000 per input, 500 inputs per cell. PGD surrogate reads are charged and are 83.2\% of its query bill.} \\
\bottomrule
\end{tabular}

\caption{Tuned configuration of the three search baselines under the shared query budget. The surrogate-query fraction is the share of PGD's queries spent on the differentiable surrogate rather than the victim (\autoref{sec:marginal}).}
\label{tab:baseline_config}
\end{table}

\begin{table}[t]
\centering
\setlength{\tabcolsep}{4pt}
\footnotesize
\begin{tabular}{@{}llp{2.6cm}l@{}}
\toprule
\multicolumn{4}{@{}l}{\itshape Shared across all three algorithms} \\
\midrule
\multicolumn{2}{@{}l}{Environment steps per run} & \multicolumn{2}{l}{$8 \times 10^4$} \\
\multicolumn{2}{@{}l}{Actor and critic} & \multicolumn{2}{l}{separate 2-layer MLPs} \\
\multicolumn{2}{@{}l}{Observation / action dim.} & \multicolumn{2}{l}{39 / 5} \\
\multicolumn{2}{@{}l}{Discount $\gamma$} & \multicolumn{2}{l}{0.99} \\
\multicolumn{2}{@{}l}{Training seeds} & \multicolumn{2}{l}{3 $(1, 2, 3)$} \\
\midrule
\multicolumn{4}{@{}l}{\itshape Per algorithm} \\
\midrule
Alg. & LR & Rollout / batch & Width \\
\midrule
PPO & $10^{-4}$          & 2048 steps, minibatch 64 & $[256, 256]$ \\
SAC & $3{\times}10^{-4}$ & batch 256, buffer $10^{6}$ & $[128, 128]$ \\
A2C & $3{\times}10^{-4}$ & 16 steps & $[256, 256]$ \\
\bottomrule
\end{tabular}
\caption{Policy architecture and per-algorithm training hyperparameters, shared across all six environments and both objectives.}
\label{tab:policy_config}
\end{table}

\section{Timing Environment}\label{timing_env}
Query counts and wall-clock seconds are measured differently, because they depend on different things. A query count is a property of the algorithm and not the machine, so the per-success query figures ($\bar{q}$, and the query-currency panels of \autoref{fig:marginal_cost} and \autoref{fig:amortization}) pool runs across the full compute cluster sweep, spread over hundreds of heterogeneous nodes. Seconds are hardware-dependent, so every wall-clock measurement instead comes from a single-threaded probe on a fixed host, which is what makes seconds-per-query a meaningful unit (\autoref{sec:cost-model}). This is why we separate the two currencies rather than mixing them.

All probes ran under an identical slot: a 16\,GiB cgroup memory cap and every threading control pinned to one thread, so a host's physical core count is irrelevant to what a probe consumes. The software stack is identical across probes: Python 3.10.20, numpy 2.2.6, scikit-learn 1.7.2, xgboost 3.2.0, torch 2.13.0, stable-baselines3 2.9.0, glibc 2.41.

\autoref{tab:timing_hosts} lists the host per probe. The marginal-cost seconds (\autoref{fig:marginal_cost}) come from a single probe, so all twelve per-success cells are mutually comparable by construction. The amortization seconds (\autoref{fig:amortization}) come from two probes, one per objective, on separate EPYC 7763 hosts under the same slot; within an objective the seconds are exact, and across the two they agree to the $3\%$ the calibration benchmark below records. A fourth probe at $\rho = 0.16$ ran on a slower Intel Xeon and supplies no result in the body.

\begin{table}[t]
\centering
\setlength{\tabcolsep}{4pt}
\footnotesize
\begin{tabular}{@{}lp{2.9cm}rr@{}}
\toprule
Probe & CPU & MHz & RAM \\
\midrule
marginal + amort.\ (evasion) & AMD EPYC 7763 & 2445 & 527 \\
\addlinespace
amortization (inflation) & AMD EPYC 7763 & 2445 & 528 \\
\addlinespace
$\rho = 0.16$ (excluded) & Intel Xeon Silver 4114 & 2200 & 528 \\
\bottomrule
\end{tabular}
\caption{Host CPU per timing probe. The two EPYC probes supply every seconds-based result in the body; the Xeon probe is excluded from cross-probe comparison (\autoref{tab:timing_calibration}). Every probe is single-threaded regardless of host core count.}
\label{tab:timing_hosts}
\end{table}

Each probe runs two fixed micro-benchmarks before measurement: a $512^2$ numpy matrix multiply and a forward pass through a width-256 MLP, the operations that dominate search and policy cost respectively. \autoref{tab:timing_calibration} reports both. The EPYC probes agree to within $1.6\%$ on the matmul and $3\%$ on the forward pass, while the Xeon is $2.3\times$ and $2.8\times$ slower, which is why its seconds are not comparable and are excluded. One residual source of variance survives this calibration: the two amortization probes disagree by roughly $30\%$ on the policy's training time $T_{\text{train}}$ (\autoref{sec:amortization}), so the seconds break-even $T^*$ is a point estimate good to roughly $1.3\times$, not to the quoted digits. The query break-even $N^*$ is unaffected, since it never touches a probe.

\begin{table}[t]
\centering
\setlength{\tabcolsep}{4pt}
\footnotesize
\begin{tabular}{@{}lrr@{}}
\toprule
Probe & Matmul (ms) & Forward (ms) \\
\midrule
marginal + amort.\ (evasion) & 7.10 & 0.566 \\
amortization (inflation) & 6.88 & 0.552 \\
$\rho = 0.16$ (excluded) & 15.97 & 1.535 \\
\midrule
\multicolumn{3}{@{}p{\dimexpr\columnwidth-2\tabcolsep}@{}}{EPYC probes agree to $1.6\%$ (matmul) and $3\%$ (forward); the Xeon is $2.3\times$ and $2.8\times$ slower.} \\
\bottomrule
\end{tabular}
\caption{Per-probe micro-benchmarks. The EPYC probes agree to within $1.6\%$ (matmul) and $3\%$ (forward pass); the Xeon is $2.3$--$2.8\times$ slower. Together with the $T_{\text{train}}$ variance noted in the text, this bounds the uncertainty on seconds-based results.}
\label{tab:timing_calibration}
\end{table}

\end{document}